\documentclass[10pt,twocolumn,twoside]{IEEEtran}

\usepackage[algoruled,vlined]{algorithm2e}
\usepackage{psfig}
\usepackage{subfigure}
\usepackage[flushleft]{threeparttable}
\usepackage{amsmath,amssymb,amsthm}
\usepackage{pgf}
\usepackage{latexsym}
\usepackage{stmaryrd}
\usepackage{graphicx}
\usepackage{color}
\usepackage{pgf}
\usepackage{multirow}
\usepackage{paralist}

\begin{document}

\title{Multiagent Conflict Resolution for a Specification Network of Discrete-Event Coordinating Agents}
%\title{Distributed Coordination Synthesis for Nonblocking Discrete-event Agents: A Specification Network Decomposition and AND/OR Graph Planning Approach}

\author{Manh Tung Pham and Kiam Tian Seow, {\em Senior Member, IEEE}}

\maketitle

\def\QED{\mbox{\rule[0pt]{1.5ex}{1.5ex}}}
\theoremstyle{plain}
\newtheorem{definition}{Definition}
\newtheorem{example}{Example}
\newtheorem{lemma}{Lemma}
\newtheorem{theorem}{Theorem}
\newtheorem{corollary}{Corollary}
\newtheorem{remark}{Remark}
\newtheorem{proposition}{Proposition}
\newtheorem{problem}{Problem}

\begin{abstract}
This paper presents a novel compositional approach to distributed
coordination module (CM) synthesis for multiple discrete-event
agents in the formal languages and automata framework. The approach
is supported by two original ideas. The first is a new formalism
called the Distributed Constraint Specification Network (DCSN) that
can comprehensibly describe the networking constraint relationships
among distributed agents. The second is multiagent conflict
resolution planning, which entails generating and using AND/OR
graphs to compactly represent conflict resolution
(synthesis-process) plans for a DCSN. Together with the framework of
local CM design developed in the authors' earlier work, the
systematic approach supports separately designing local and
deconflicting CM's for individual agents in accordance to a selected
conflict resolution plan. Composing the agent models and the CM's
designed furnishes an overall nonblocking coordination solution that
meets the set of inter-agent constraints specified in a given DCSN.
\end{abstract}

\section{Introduction}
In the paradigm of discrete-event systems (DES's), one can distinguish two fundamental types of control to satisfy given specifications. Specifications prescribe constraints that assert some orderly flow of system activities based on system needs or limitations \cite{J-Seow2009}. One type is that of external supervisors controlling discrete-event processes or agents to satisfy given control constraints \cite{J-PeterRamadge1987}, while the other type is that of agents coordinating among themselves through their coordination modules (CM's) to satisfy given inter-agent constraints \cite{J-Pham2011,J-Seow2009,CAAMAS-Seow2004}. The CM's are built-in strategies designed for the given constraints, and constitute an agent's local interface ``plugged'' onto the agent model via the synchronization operator, and through which every agent coordinates by interacting and communicating with other agents in the system. Although the two types of control are mathematically related, they are clearly conceptually different \cite{J-Seow2009}.

In \cite{J-Pham2011,J-Seow2009,CAAMAS-Seow2004}, we formulate and address the fundamental coordination problem of multiple agents coordinating to satisfy one common constraint. Therein, by establishing the mathematical connection between the discrete-event coordination problem with the conceptually different discrete-event supervisory control problem \cite{J-PeterRamadge1987}, we successfully adapt concepts and techniques from supervisory control of DES's \cite{J-PeterRamadge1987} for the development of a CM synthesis algorithm.

In this paper, we generalize the fundamental coordination problem \cite{J-Pham2011} to a networked coordination problem of multiple agents coordinating to satisfy multiple constraints distributed among them. The inter-agent constraints are distributed in such a way that each constraint is pre-specified for a subgroup of agents. These agent subgroups can be overlapping, meaning that an agent can be coordinating on different inter-agent constraints with different agents in the system, and hence conflict or blocking between their different coordinating actions may arise. In general, multiagent conflict can occur if some agent actions in a system state can permanently prevent some of the agents in the system from reaching their local design goals characterized by marked states under the discrete-event paradigm. This presents a challenging design problem of networked agent coordination which is commonly encountered in large scale distributed systems.

To address the networked coordination problem for large scale DES's in a systematic fashion, we propose a novel compositional synthesis approach. This approach consists of two main steps. In the first step, we construct for each agent a set of local CM's, one for each of the agent's relevant constraints, using the synthesis algorithm proposed in \cite{J-Pham2011}. The advantage of constructing local CM's is that we can avoid having to compute the product of all agent and constraint models, thereby mitigating the problem of state explosion. In the second step, we generate a conflict resolution plan, and execute this plan to design additional deconflicting CM's for individual agents. A conflict resolution plan for a DCSN shows a sequential or partial order of applying deconflicting CM synthesis to successive pairs of potentially conflicting, constrained agent subgroups. Deconflicting CM's are individual agent CM's to be interposed between every agent model and its local CM's, so that in coordinating among themselves, the agents can automatically resolve the conflicts that may otherwise occur due to the different inter-agent constraints on which each agent's local CM's are synthesized. Composing the agent models and the local and deconflicting CM's can then be shown to constitute a correct solution to the networked coordination problem.

Our compositional synthesis approach for designing distributed coordinating agents is supported by two original ideas. The first is a new formalism called the Distributed Constraint Specification Network (DCSN) that can describe comprehensibly the networking constraint relationships among agents, on which the multiagent networked coordination problem is formulated and addressed (Sections \ref{Sect-ProblemFormulation} and \ref{Sect-SubNetSynthesis}). The second is multiagent conflict resolution planning that entails generating a compact AND/OR graph representation \cite{B-NilsNilsson1980} of conflict resolution plans and selecting some criterion-based optimal plan for a given DCSN (Section \ref{Sect-ConflictResolutionPlanning}). At the outset, the background and preliminaries are presented (Section \ref{Sect-Pre}). An example system introduced in Section \ref{Sect-ProblemFormulation} is used throughout the paper to illustrate the various aspects of the proposed approach. The proofs of all new results are presented in the appendix. A summary and a discussion of related work conclude the paper (Section \ref{Sect-RelatedWork-Conclusion}).

\section{Background and Preliminaries} \label{Sect-Pre}
In this paper, we will use small letters such as $n$, $m$, $k$, $r$ to denote integers. For an integer $n \geq 1$, the symbol $I_n$ denotes the index set $\{1, 2, ... , n \}$.

\subsection{Languages and Automata}
Let $\Sigma$ be a finite alphabet of symbols representing individual events. A {\em string} is a finite sequence of events from $\Sigma$. Denote $\Sigma^*$ as the set of all strings from $\Sigma$ including the empty string $\varepsilon$. A string $s^\prime$ is a {\em prefix} of $s$ if $(\exists t \in \Sigma^*)$ $s^{\prime}t = s$. A {\em language} $L$ over $\Sigma$ is a subset of $\Sigma^*$. Say $L_1$ is a {\em sublanguage} of $L_2$ if $L_1 \subseteq L_2$. The {\em prefix closure} $\bar{L}$ of a language $L$ is the language consisting of all prefixes of its strings. Clearly $L \subseteq \bar{L}$.%, because any string $s$ in $\Sigma^*$ is a prefix of itself.
 A language $L$ is {\em prefixed-closed} if $L = \bar{L}$.

Given $\Sigma^1 \subseteq \Sigma^2$, the natural projection $P_{\Sigma^2, \Sigma^1}: (\Sigma^2)^* \rightarrow (\Sigma^1)^*$, which erases from a string $s \in (\Sigma^2)^*$ every event $\sigma \in (\Sigma^2 - \Sigma^1)$, is defined recursively as follows: $P_{\Sigma^2, \Sigma^1} (\varepsilon) =
 \varepsilon$, and $(\forall s \in (\Sigma^2)^*)(\forall \sigma \in \Sigma^2)$, $P_{\Sigma^2, \Sigma^1} (s\sigma) = P_{\Sigma^2, \Sigma^1} (s)\sigma$, if $\sigma \in \Sigma^1$, and $P_{\Sigma^2, \Sigma^1} (s)$, otherwise.

For $L \subseteq (\Sigma^2)^*$, $P_{\Sigma^2, \Sigma^1}(L) \subseteq (\Sigma^1)^*$ denotes the language $\{ P_{\Sigma^2, \Sigma^1}(s) \mid s \in L \}$. The inverse image of $P_{\Sigma^2, \Sigma^1}$, denoted by $P^{-1}_{\Sigma^2, \Sigma^1}$, is a mapping from $(\Sigma^1)^*$ to $(\Sigma^2)^*$, and defined as: for $L_1 \in (\Sigma^1)^*$, $P^{-1}_{\Sigma^2, \Sigma^1}(L_1) = \{ L \subseteq (\Sigma^2)^* \mid P_{\Sigma^2, \Sigma^1}(L) = L_1\}$. Clearly, for $L \in (\Sigma^2)^*$, $P_{\Sigma^2, \Sigma^1}^{-1}(P_{\Sigma^2, \Sigma^1}(L)) \supseteq L$.

If a language is {\em regular} \cite{L-Wonham}, then it can be {\em generated} by an automaton. An {\em automaton} $A$ is a 5-tuple $(X^A, \Sigma^A, \delta^A, x^A_0, X^A_m)$, where $X^A$ is the finite set of states, $\Sigma^A$ is the finite set of events, $\delta^A: \Sigma^A \times X^A \rightarrow X^A$ is the (partial) transition function, $x^A_0$ is the {\em initial state} and $X^A_m \subseteq X^A$ is the subset of {\em marker states}.

The definition of $\delta^A$ can be extended to $(\Sigma^A)^* \times X^A$ as follows: $\delta^A (\varepsilon, x) = x$,
and $(\forall \sigma \in \Sigma^A)(\forall s \in (\Sigma^A)^*)\delta^A (s\sigma, x) = \delta^A(\sigma, \delta^A(s,x))$. Write $\delta^A(\sigma,x)!$ to denote that $\delta^A(\sigma,x)$ is defined. The behaviors of automaton $A$ can then be described by the prefix-closed language $L(A)$ and the marked language $L_m(A)$. Formally, $L(A) = \{ s \in (\Sigma^A)^* \mid \delta^A (s, x_0)! \}$, and $L_m (A) = \{ s \in L(A) \mid \delta^A (s, x_0) \in X^A_m \}$.

Let $A_i$, $i \in \{ 1,2 \}$, be two automata. Then their {\em synchronous product} $A$, denoted by $A = A_1\parallel A_2$, models a discrete-event system (DES) of $A_1$ and $A_2$ operating concurrently by interleaving events generated by $A_1$ and $A_2$, with synchronization on shared events $\sigma \in \Sigma^{A_1}\cap \Sigma^{A_2}$. It has been shown that if $A = A_1 \parallel A_2$ then $L(A) = P^{-1}_{\Sigma^A, \Sigma^{A_1}}(L(A_1)) \cap P^{-1}_{\Sigma^A, \Sigma^{A_2}}(L(A_2))$ and $L_m(A) = P^{-1}_{\Sigma^A, \Sigma^{A_1}}(L_m(A_1)) \cap P^{-1}_{\Sigma^A, \Sigma^{A_2}}(L_m(A_2))$ \cite{L-Wonham}. If $\Sigma^{A_1} = \Sigma^{A_2}$, then $L(A_1 \parallel A_2) = L(A_1) \cap L(A_2)$ and $L_m(A_1 \parallel A_2) = L_m(A_1) \cap L_m(A_2)$. The synchronous product of $n \geq 2$ automata $A_1$, $A_2$, ... $A_n$, denoted by $\parallel^{i=n}_{i=1} A_i$, can be defined recursively using the associativity of $\parallel$ \cite{L-Wonham}.

\subsection{Nonblocking Coordination among Discrete-event Agents}
Let $\mathcal{A} = \{ A_i \mid i \in I_n \}$ be a set of $n \geq 2$ nonblocking automata modeling $n$ discrete-event agents, with $\Sigma^{A_i} \cap \Sigma^{A_j} = \emptyset$ for $i \neq j$.  The event set $\Sigma^{A_i}$ (of agent $A_i$) is partitioned into the controllable event set $\Sigma^{A_i}_{c}$ and  the uncontrollable event set $\Sigma^{A_i}_{uc}$. Interpreted from the agent viewpoint, an uncontrollable event is inherently autonomous and can be executed solely at the free will of the owner agent.

Let $A = A_1 \parallel A_2 \parallel ... \parallel A_n$ model a system of $n$ agents in $\mathcal{A}$ freely interacting, with $\Sigma^A_c = \underset{i \in I_n}{\bigcup}\Sigma^{A_i}_c$ and $\Sigma^A_{uc} = \underset{i \in I_n}{\bigcup}\Sigma^{A_i}_{uc}$. Let $J \subseteq I_n$. Then, an inter-agent constraint for a group of agents $\mathcal{A}_J = \{A_j \mid j \in J \}$ can be prescribed by an automaton $C_J$ such that $(\forall j \in J)\Sigma^{C_J} \cap \Sigma^{A_j} \neq \emptyset$. The language $L_m(C_J)$ is interpreted as the set of desirable event sequences that one wishes to impose on the group of agents $\mathcal{A}_J$. In other words, constraint $C_J$ specifies that the agents in $\mathcal{A}_J$ must coordinate among themselves so that none of those event sequences in $L_m(A_J)-L_m(A_J \parallel C_J)$ will ever be generated during their interaction, where $A_J = \underset{j \in J}{\parallel} A_j$. $C_J$ is then said to be a relevant constraint for agent group $\mathcal{A}_{J}$.

\begin{definition} {\em \cite{J-Pham2011}:} \label{def-CoordinationModule} A coordination module (CM) for an agent $A_i$, $i \in I_n$, is an automaton $S^h_i$ with the following properties: (i) $\Sigma^{A_i} \subseteq \Sigma^{S^h_i}$, and (ii) $S^h_i$ is $(\Sigma^{S^h_i} - \Sigma^{A_i}_c)$-enabling, namely, $(\forall s \in (\Sigma^A)^*)(\forall \sigma \in (\Sigma^{S^h_i} - \Sigma^{A_i}_{c}))$ $[(s \in L(S^h_i \parallel A)\ \mbox{ and } s\sigma \in L(A)) \Rightarrow s\sigma \in L(S^h_i \parallel A)]$.
\end{definition}

Through their CM's, the agents coordinate as follows. Following the execution of a string $s \in L(A)$, $A_i$ updates the state of every CM $S^h_i$ to $x^h_i = \delta^{S^h_i}(P_{\Sigma^A, \Sigma^{S^h_i}}(s), x^{S^h_i}_0)$. $A_i$ then enables (allows to execute) only events $\sigma_i \in \Sigma^{A_i}$ that is defined at every current state of its CM's. The result is that the system behavior is restricted to a sublanguage of $L(A)$.

That each CM $S^h_i$ is $(\Sigma^{S^h_i} - \Sigma^{A_i}_c)$-enabling guarantees that $A_i$ only disables its own controllable events. In other words, $A_i$ always enables (and hence never prevents from execution) its uncontrollable events and never interferes with the execution of events of the other agents. $\Sigma^{S^h_i}$ represents the set of events that $A_i$ needs to observe in order to correctly update the state of $S^h_i$ when interacting with the other agents. The event set $(\Sigma^{S^h_i} - \Sigma^{A_i})$, which cannot be observed locally by $A_i$, must be communicated to $A_i$ by other agents.

Let $\mathcal{CM} = \{\mathcal{CM}_i \mid i \in I_n \}$ and $CM_i = \underset{S^h_i \in \mathcal{CM}_i}{\parallel} S^h_i$. The system of $n$ agents in $\mathcal{A}$ coordinating through their respective CM's can then be represented by $A^{CM}= \underset{i \in I_n}{\parallel}(A_i \parallel CM_i)$. The CM's are then said to be nonblocking if every string generated during the agents' interaction can be completed to a marked string, i.e., $\overline{L_m(A^{CM})} = L(A^{CM})$.

The fundamental problem of multiple agents coordinating to respect one constraint may now be stated as follows: Given $n$ agents $A_i$, $1 \leq i \leq n$, and an inter-agent constraint $C$, construct a nonblocking CM set $\{S_i \mid 1 \leq i \leq n\}$, where $S_i$ is for $A_i$, such that $L_m(\parallel^{n}_{i=1} (A_i || S_i))$ is equal to the supremal controllable sublanguage \cite{J-PeterRamadge1987} of $L_m(A) \cap L_m(C)$.

Theorem \ref{Th-MC-Pro-1} addresses the fundamental problem of multiple agents coordinating to respect one constraint. It is expressed in terms of the concepts of language controllability (Definition \ref{def-ControllabelLanguage}) and language observability (Definition \ref{def-ObservableLanguage}).

\begin{theorem} \label{Th-MC-Pro-1}
Given $n \geq 2$ agent automata $A_i$, $1 \leq i \leq n$, with $\Sigma^{A_i} \cap \Sigma^{A_j} = \emptyset$ for $i \neq j$. Let $A = \parallel^{n}_{i=1} A_i$, $\emptyset \neq K \subseteq L_m(A)$ and $\Sigma_{com} \subseteq \Sigma^A$. Then, there exists a CM set $\{ S_i \mid 1 \leq i \leq n \}$, where $S_i$ is for $A_i$, such that  $L_m(\parallel^{n}_{i=1} (A_i || S_i)) = K$,  $L(\parallel^{n}_{i=1} (A_i || S_i)) = \bar{K}$  and $\bigcup^{n}_{i=1} (\Sigma^{S_i} - \Sigma^{A_i}) = \Sigma_{com}$, if and only if $K$ is coordinable w.r.t $A$ and $\Sigma_{com}$, namely, $K$ is controllable w.r.t $A$ and $\Sigma^A_c = \bigcup^{n}_{i=1}$ and
$K$ is observable w.r.t $A$ and $P_{\Sigma^A, \Sigma^{A_i} \cup \Sigma_{com}}$ for all $1 \leq i \leq n$.
\end{theorem}

Theorem \ref{Th-MC-Pro-1} follows from the fact that supervision and multiagent coordination are mathematically equivalent, as established and discussed in \cite{CAAMAS-Seow2004}. Importantly, in Theorem \ref{Th-MC-Pro-1}, $\Sigma_{com}$ constitutes the system communication set, which is a union of local event subsets to be communicated to each agent. As explained in \cite{J-Pham2011}, unlike supervisory control, the observable events for a receiving agent (or events to be communicated to the agent when they occur) are not pre-determined but computed with the aim of minimizing communication, and therefore can be different for a different inter-agent constraint.

\begin{definition}{\em \cite{J-PeterRamadge1987}: }\label{def-ControllabelLanguage}
$K \subseteq L(A)$ is said to be {\em controllable} with respect to (w.r.t) $A$ and $\Sigma^A_c$ (or just controllable if $\Sigma^A_c$ is understood) if $(\forall s \in \overline{K})(\forall \sigma \in \Sigma^A_{uc})$ $[s\sigma \in L(A) \Rightarrow s\sigma \in \overline{K}]$.
\end{definition}

In other words, $K$ is controllable provided no $L(A)$-string which is already a prefix of some string in $K$, that when followed by an uncontrollable event in $\Sigma^A_{uc}$, would exit from $\overline{K}$. It has been shown that the {\em supremal controllable sublanguage} \cite{J-PeterRamadge1987} of $K$ w.r.t $A$ and $\Sigma^A_c$ exists, and is equal to $K$ if it is controllable. For an automaton $C$, the $Supcon(C, A, \Sigma^A_c)$ procedure \cite{SW-TCT}, which computes a nonblocking automaton $S$ such that $L_m(S)$ is the supremal controllable sublanguage of $L_m(A) \cap L_m(C)$, can be implemented with polynomial time complexity \cite{L-Wonham}.

\begin{definition}{\em \cite{J-FengLin1988}: }\label{def-ObservableLanguage}
$K \subseteq L_m(A)$ is said to be {\em observable}  w.r.t $A$ and
$P_{\Sigma^A, \Sigma^A_o}$ (or just observable if $P_{\Sigma^A, \Sigma^A_o}$ is understood) if $(\forall s, s^\prime \in (\Sigma^A)^*)$ for which $P_{\Sigma^A, \Sigma^A_o}(s) = P_{\Sigma^A, \Sigma^A_o}(s^\prime)$, the following two conditions are satisfied: (1) $(\forall \sigma \in \Sigma^A)[(s\sigma \in \overline{K} \mbox{ and } s^\prime \in \overline{K} \mbox{ and } s^\prime\sigma \in L(A)) \Rightarrow s^\prime\sigma \in \overline{K}]$, and (2) $[s \in K \mbox{ and }s^\prime \in \overline{K} \cap L_m(A)] \Rightarrow s^\prime \in K$.
\end{definition}

The above conditions ensure that $\Sigma^A_o$ provides a sufficient view for an observer to determine all necessary control and marking actions. Taken together, that $K$ is coordinable w.r.t $A$ and $\Sigma_{com}$ means that (i) if each agent coordinates properly (by appropriately enabling and disabling its own controllable events), then the coordinated system behavior will conform to $K$, and (ii) $A_i$ has sufficient information for determining its coordinating actions (that ensure the conformance of the coordinated system behavior to $K$).

\section{Problem Formulation} \label{Sect-ProblemFormulation}

\subsection{Distributed Constraint Specification Network}
In distributed multiagent systems, there are often multiple distributed inter-agent constraints, each restricting a group of interacting agents. To specify the relevance relationships of distributed constraints among these agents, we define a formalism called the distributed constraint specification network (DCSN). The DCSN allows a human designer to organize and interconnect the agents and their distributed constraints in a networking structure that, in our opinion, comprehensibly shows \lq\lq who needs to coordinate with whom over what constraints''.

\begin{definition} \label{Ch-NetworkedPartI-Def-DCSN}
Let $n \geq 2$, $m \geq 1$.  A distributed constraint specification network (DCSN) $\mathcal{N}$ is a tuple $(\mathcal {A}, \mathcal {C})$, where $\mathcal {A} = \{ A_i \mid i \in I_n\}$ is an agent set of size $n$ and  $\mathcal {C} = \{ C^k_{J_k} \mid k \in I_m, J_k \subseteq I_n \}$ is an inter-agent constraint set of size $m$, such that $(\forall C^k_{J_k} \in \mathcal{C})(\forall i \in J_k) \Sigma^{A_i}\cap \Sigma^{C^k_{J_k}} \neq \emptyset$.%\hspace{\fill}$\blacksquare$
\end{definition}

Each $C^k_{J_k} \in \mathcal{C}$ in a DCSN $\mathcal{N}$, where $k$ is the constraint index, is said to be a relevant constraint for agents in the group $\mathcal{A}_{J_k} = \{A_i \mid i \in J_k\}$. Without loss of generality, assume henceforth that $\underset{k \in I_m}{\bigcup} J_k = I_n$, i.e., every agent in $\mathcal{A}$ is in $\mathcal{A}_{J_k}$ for some $k$, and so every agent needs to coordinate. Then a DCSN can be redefined as $\mathcal{N} = \{ (J_k, C^k_{J_k}) \mid k \in I_m, J_k \subseteq I_n\}$.

\begin{definition} \label{Ch-NetworkedPartI-Def-Subnet}
An element $\mathcal{N}^k_1 = (J_k, C^k_{J_k})$ of $\mathcal{N}$ is called a basic subnet of  $\mathcal{N}$; and a non-empty $\mathcal{N}^{\mathcal{S}_r}_r \subseteq \mathcal{N}$ consisting of $r = |\mathcal{S}_r| \geq 1$ basic subnets is called a $r$-constraint subnet of $\mathcal{N}$ with constraint subset $\{C^k_{J_k} \mid k \in \mathcal{S}_r\}$. Where the constraint subset is arbitrary, a $r$-constraint subnet is simply denoted by $\mathcal{N}_r$.%\hspace{\fill}$\blacksquare$
\end{definition}

By Definition \ref{Ch-NetworkedPartI-Def-Subnet}, a subnet of a DCSN is also a DCSN. Intuitively, a DCSN is a formalism that represents interconnections among agents and constraints, associating every agent with its relevant inter-agent constraints. Under the interconnections, an inter-agent constraint induces a group of agents that it is relevant for. It is then clear that the agents in the agent group need to coordinate to satisfy the constraint.

A DCSN can be graphically represented by an undirected hyper-graph with agents represented by rectangular nodes, and each constraint relevant for an agent group by an oval hyper-edge with arcs connecting it to all the agents in the group. Through its graphical representation which is intuitively clear and easy to understand, a DCSN is designer comprehensible for modeling the inter-agent constraint relationships among agents, as the following example will demonstrate.

\begin{example}
Throughout this paper, we shall use a simple manufacturing transfer line example [Fig. \ref{Ch-NetworkedPartI-Fig-Overall}] to illustrate our theoretical development. The system under study consists of three agents $A_1$, $A_2$ and $A_3$ [Figs. \ref{Ch-NetworkedPartI-Fig-A1}--\ref{Ch-NetworkedPartI-Fig-A3}], and four constraints $E^1_{\{1,2\}}$, $E^2_{\{1,2\}}$, $B^3_{\{1,3\}}$ and $B^4_{\{2,3\}}$ [Figs. \ref{Ch-NetworkedPartI-Fig-E1}--\ref{Ch-NetworkedPartI-Fig-B2}], organized into a DCSN (Fig. \ref{Ch-NetworkedPartI-Fig-NW}).

The system works as follows. $A_1$ and $A_2$ are producer agents that continually follow a production plan: Acquire manufacturing equipment $E_1$ and $E_2$ in either order, produce a workpiece, return the equipment to their initial location, move to the buffers' location, place the finished workpiece into the respective buffer, and finally return to the initial state for a new production cycle. $A_3$ is a delivery agent that continually takes a work piece from either buffer 1 or buffer 2, processes, and delivers it to customers. We fix $\Sigma^A_{uc} = \{1produce, 1return, 1place, 2produce, 2return, 2place,$ $3process, 3deliver \}$. The four constraints $E^1_{\{1,2\}}$, $E^2_{\{1,2\}}$, $B^3_{\{1,3\}}$ and $B^4_{\{2,3\}}$ are formulated to respectively ensure mutual exclusion of equipment use, and no overflow or underflow of buffers.

\begin{figure}[ttt]
\centering
\subfigure[Overall system model]
{
    \psfig{file=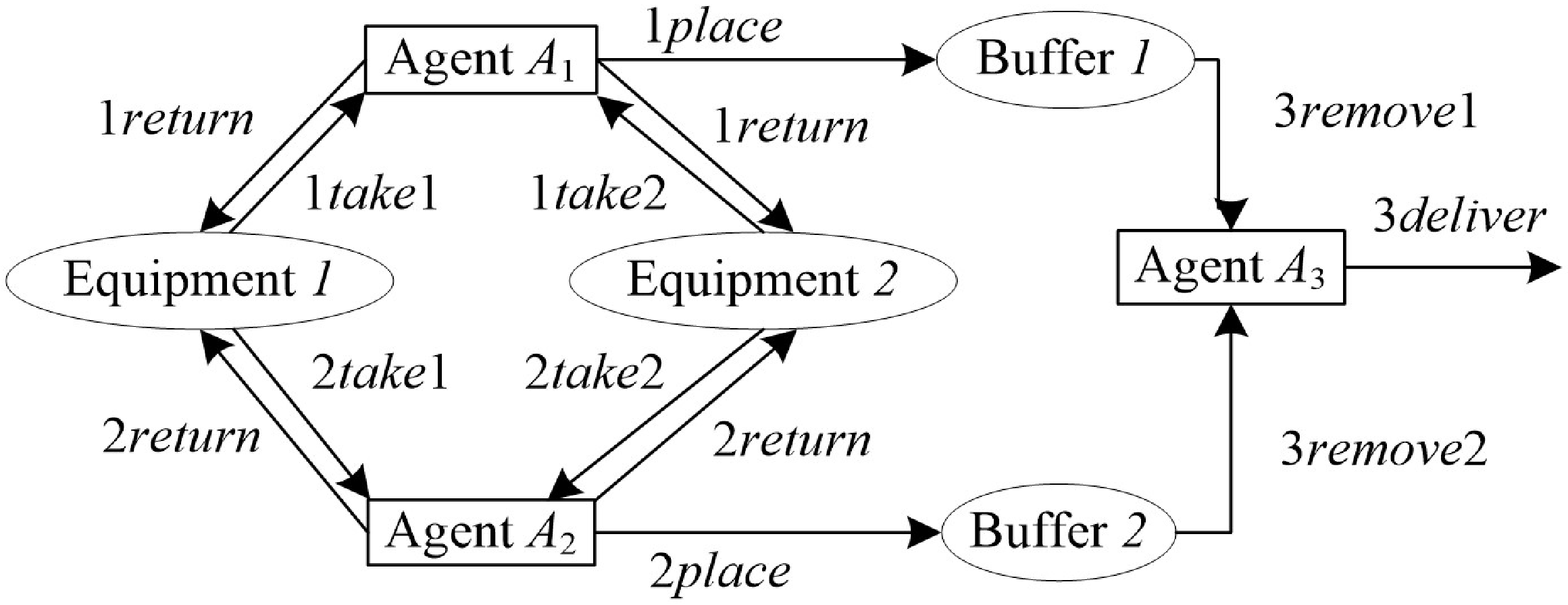,width=7cm} \label{Ch-NetworkedPartI-Fig-Overall}
}
\subfigure[Agent $A_1$]
{
    \psfig{file=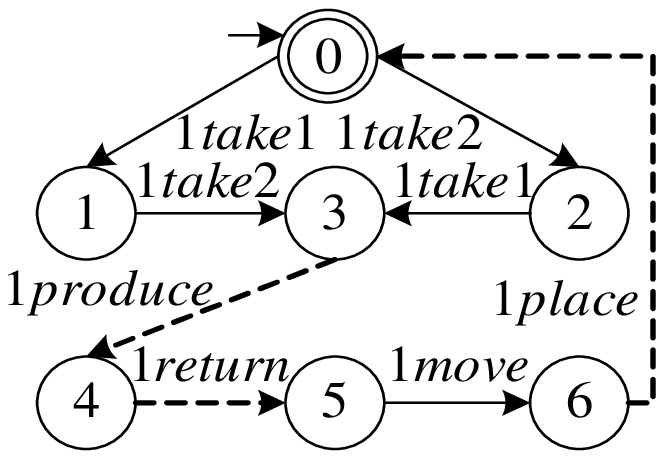,width=3cm} \label{Ch-NetworkedPartI-Fig-A1}
}
\subfigure[Agent $A_2$]
{
    \psfig{file=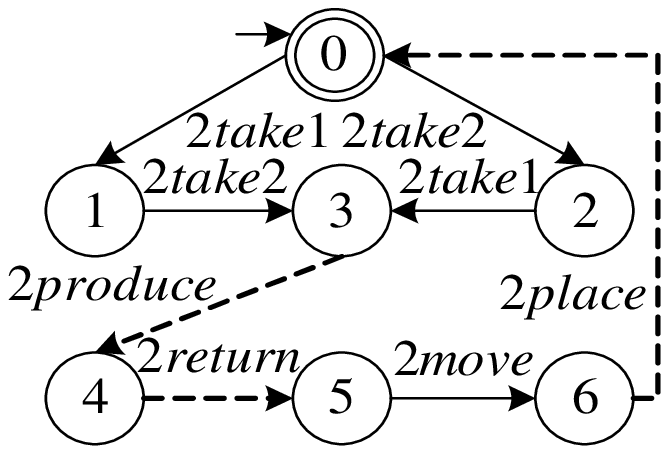,width=3cm} \label{Ch-NetworkedPartI-Fig-A2}
}
\subfigure[Agent $A_3$]{
        \psfig{file=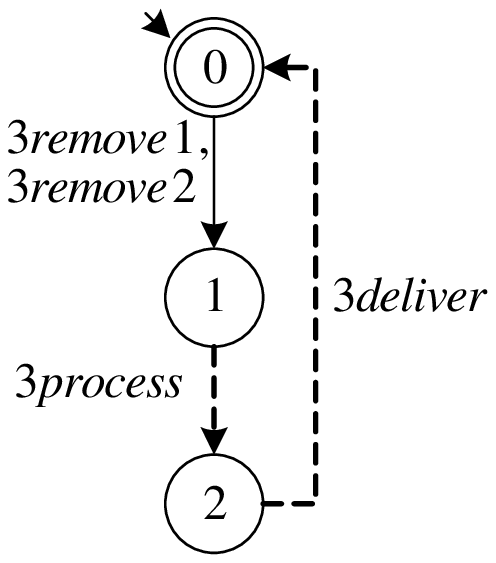,width=1.8cm} \label{Ch-NetworkedPartI-Fig-A3}
}

\subfigure[$E^1_{\{1,2\}}$]
{
    \psfig{file=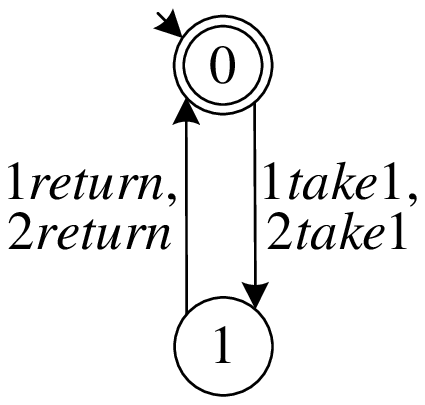,width=1.8cm} \label{Ch-NetworkedPartI-Fig-E1}
}
\subfigure[$E^2_{\{1,2\}}$]
{
    \psfig{file=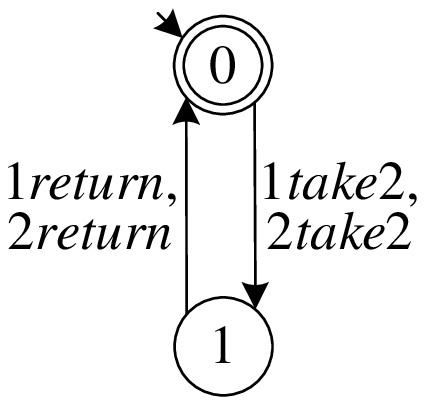,width=1.8cm} \label{Ch-NetworkedPartI-Fig-E2}
}
\subfigure[$B^3_{\{1,3\}}$]{
        \psfig{file=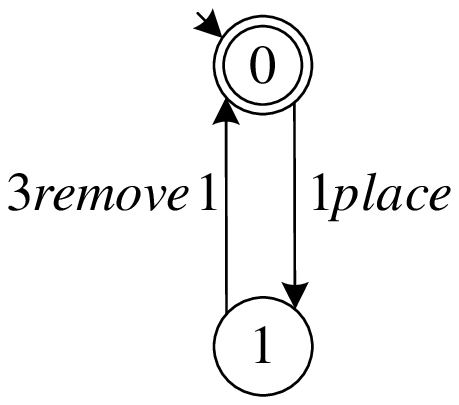,width=1.8cm} \label{Ch-NetworkedPartI-Fig-B1}
}
\subfigure[$B^4_{\{2,3\}}$]{
        \psfig{file=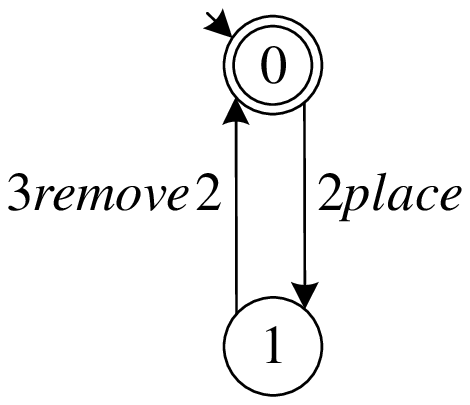,width=1.8cm} \label{Ch-NetworkedPartI-Fig-B2}
}
\subfigure[DCSN]{
   \psfig{file=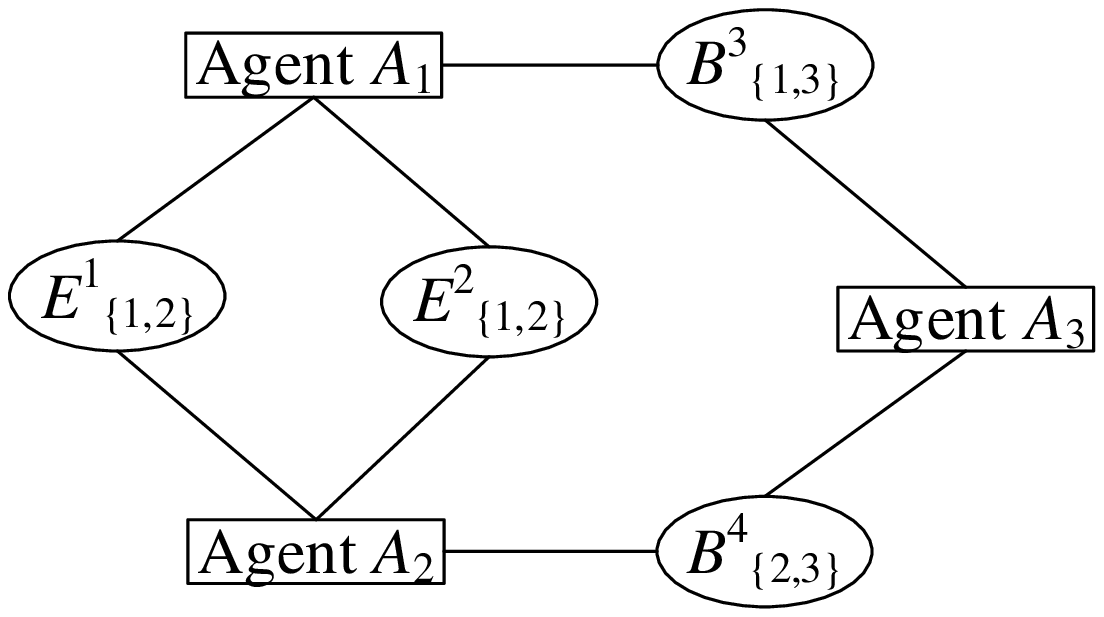,width=4cm} \label{Ch-NetworkedPartI-Fig-NW}
}
\caption{A manufacturing transfer line system.} \label{Ch-NetworkedPartI-Fig-ManfacturingEx}
\end{figure}

The DCSN  is composed of four basic subnets $\mathcal{N}^1_1 = (\{1,2\}, E^1_{\{1,2\}})$, $\mathcal{N}^2_1 = (\{1,2\}, E^2_{\{1,2\}})$, $\mathcal{N}^3_1 = (\{1,3\}, B^3_{\{1,3\}})$ and $\mathcal{N}^4_1 = (\{2,3\}, B^4_{\{2,3\}})$. When depicted graphically, a nice feature of DCSN is that the constraint inter-connections between agents are explicitly shown for comprehensibility of design. For instance, in Fig. \ref{Ch-NetworkedPartI-Fig-NW}, it is clear that $A_1$ would need to coordinate with $A_2$ for $E^1_{\{1,2\}}$ and $E^2_{\{1,2\}}$, and with $A_3$ for $B^3_{\{1,3\}}$.
\end{example}

\subsection{Networked Coordination Problem Statement}
%We can now formally state our discrete-event networked coordination problem.

\begin{problem} \label{Ch-NetworkedPartI-Prob-Main}
Given a DCSN $\mathcal{N} = (\mathcal{A}, \mathcal{C})$ of $n$ agents and $m$ inter-agent constraints, let $A = \underset{i \in I_n}{\parallel} A_i$ and $C = \underset{k \in I_m}{\parallel} C^k_{J_k}$, where $A_i \in \mathcal{A}$ and $C^k_{J_k} \in \mathcal{C}$. Synthesize a set $\mathcal{CM} = \{ \mathcal{CM}_i \mid i \in I_n \}$, where $\mathcal{CM}_i$ is a set of CM's for agent $A_i$, such that $A^{CM} \equiv Supcon(C,A)$, i.e., the resulting coordinated system is nonblocking and satisfies every constraint in $\mathcal{C}$ in a minimally restrictive manner.%\hspace{\fill}$\blacksquare$
\end{problem}

$L_m(C)$ specifies the desired behavior, embodying all the event sequences that one wishes to impose on the system $A$. A set $\mathcal{CM}$ of CM's is then said to satisfy (every constraint in) $\mathcal{C}$ if $L_m(A^{CM}) \subseteq L_m(C)$. It can be easily shown that $L_m(A^{CM})$ is controllable with respect to $A$ and $\Sigma^A_{uc}$. Thus, for a set $\mathcal{CM}$ of CM's satisfying $\mathcal{C}$, $L_m(A^{CM}) \subseteq L_m(Supcon(C,A))$. A CM set $\mathcal{CM}$ is then said to satisfy $\mathcal{C}$ in a {\em minimally interventive} manner if  $A^{CM} \equiv Supcon(C,A)$, implying that using such CM's, each agent $A_i$ would not unnecessarily disable its controllable events, unless not doing so could lead eventually to the violation of some inter-agent constraint in $\mathcal{C}$.

\subsection{Compositional Synthesis} \label{Sect-Synthesis}
As discussed in the introduction, our compositional synthesis approach for a given DCSN can be described as follows.

\textbf{- Step 1 Basic Subnet Synthesis:} Synthesize for every agent a set of $\parallel$-connected local CM's, one for each of the agent's relevant constraints. This step is performed by applying the algorithm developed in \cite{J-Pham2011} to every basic constraint subnet of the DCSN, i.e., every subnet containing one inter-agent constraint.

\textbf{- Step 2 Subnet Composition}
    \begin{itemize}
        \item \textbf{Step 2.1 Conflict Resolution Plan Generation:} Generate a conflict resolution plan for the DCSN. This plan is a sequence of subnet composition operations. Each operation entails designing deconflicting CM's for the agents of the subnets concerned, so as to ensure nonblockingness, and hence correctness, when the subnets are composed together.

        \item \textbf{Step 2.2 Conflict Resolution Plan Execution:} Compose subnets with conflict resolution by following a precedence order of subnet composition operations in the plan generated in Step 2.1. This is to completely deconflict the local CM's synthesized in Step 1 to ensure nonblockingness of the whole DCSN.
    \end{itemize}

In the remaining of this paper, we explain how these steps are formally carried out.

\section{Subnet Synthesis} \label{Sect-SubNetSynthesis}
This section fills in the CM synthesis details of our approach, presenting for Step 1, the local CM synthesis algorithm developed in \cite{J-Pham2011}, and for Step 2.2, how the CM solutions obtained of smaller subnets can be composed to obtain a nonblocking solution for the resultant bigger subnet.

Note that, having pointed out in \cite{CAAMAS-Seow2004,J-Pham2011} the mathematical relation between multiagent coordination and supervisory control, we are able to identify and utilize some mathematical results developed for supervisory control to support subnet composition synthesis, by carefully redefining these results in the notation of our DES multiagent coordination framework. In the following, the supporting results are Proposition 1, Lemma 1 and Lemma 2, and in the spirit of scientific rigor, these are validated by proofs presented in \cite{J-Pham2011} under our framework notation. In the increasingly cross-disciplinary research environment, we find it necessary to adopt this approach, in order to develop a standalone treatment of our new distributed agent coordination theory that contributes conceptually clear DES methods for multiagent coordination, without the distracting shadow of terminology from the mathematically related, but conceptually different field of supervisory control.

\subsection{Basic Subnet Synthesis}
Given a DCSN $\mathcal{N} = (\mathcal{A}, \mathcal{C})$ of $n$ agents and $m$ inter-agent constraints, we consider the problem of synthesizing CM's for some basic subnet $\mathcal{N}^k_1 = (J_k, C^k_{J_k})$ of $\mathcal{N}$, $k \in I_m$. To fix notation, let $A_{J_k} = \underset{i \in J_k}{\parallel} A_i$ and $SUP^k = Supcon(C^k_{J_k}, A_{J_k})$. We are interested in synthesizing, for each agent $A_i$ in the subnet, a CM $S^k_i$ such that $\underset{i \in J_k}{\parallel} (A_i \parallel S^k_i) \equiv SUP^k$.

The pseudo-code of the synthesis algorithm \cite{J-Pham2011} based on Theorem \ref{Th-MC-Pro-1} is notationally redefined as Procedure $CMBasicSubnet$ for basic subnet synthesis.

\DontPrintSemicolon
\begin{procedure}
\small{
\textbf{Output:} A CM $S^k_i$ for every agent $A_i$ in $\mathcal{N}^k_1=(J_k, C^k_{J_k})$

\Begin{
\textbf{Step 1}: $A_{J_k} \leftarrow \underset{i \in J_k}{\parallel} A_i, SUP^k \leftarrow Supcon(C^k_{J_k}, A_{J_k})$; \;

\textbf{Step 2}: $(\forall i \in J_k)\Sigma^k_{mincom,i} \leftarrow \Sigma^{A_i} \cup MinSysComSet(L_m(SUP^k), A_{J_K})$;\;

\textbf{Step 3}: $(\forall i \in J_k)S^k_i \leftarrow CM(SUP^k, \Sigma^k_{mincom,i})$; \;

\textbf{Step 4}: $(\forall i \in J_k)S^k_i  \leftarrow CMreduce(S^k_i, A_i)$;\;
}
\caption{$CMBasicSubnet$($\mathcal{N}^k_1$) \label{Ch-NetworkedPartI-Procedure-MCSA}}
}
\end{procedure}

Recall from \cite{J-Pham2011} that $MinSysComSet(L_m(SUP^k), A_{J_K})$ computes and returns a minimal cardinality communication event set that the agents $A_i$'s in the subnets must communicate among themselves, $CM$ constructs for each agent $A_i$, $i \in J_k$, a CM $S^k_i$ from $SUP^k$ and $\Sigma^k_{mincom,i}$, and $CMreduce$ is a CM reduced procedure adapted from the supervisor reduction procedure \cite{J-RongSu2004}, which can often return a greatly state-size reduced CM automaton for agent $A_i$, achieving the same behavior of $A_i \parallel S^k_i$.

\begin{example} \label{Ch-NetworkedPartI-Ex-Basicnet}
To illustrate the use of Procedure $CMBasicSubnet$, we apply it to the manufacturing transfer line example and synthesize CM's for agents $A_1$ and $A_2$ to cooperatively satisfy $E^1_{\{1,2\}}$. By Step 1 of $CMBasicSubnet$, we first compute $SUP^1 = Supcon(E^1_{\{1,2\}}, A_1 \parallel A_2)$, which has 40 states and 82 transitions. Next, by Step 2, the minimal communication sets for $A_1$ and $A_2$ are computed: $\Sigma^1_{mincom,1} = \{2take1, 2return\}$ and $\Sigma^1_{mincom,2}=\{1take1, 1return\}$. Following Step 3, CM's $S^1_i$, $i \in \{1, 2\}$, are computed by applying Procedure $CM$ on $SUP^i$ and $\Sigma^{A_i} \cup \Sigma^1_{mincom,i}$. Each of these CM's has 11 states and 19 transitions. Finally, in Step 4, $CMreduce$ is applied to reduce the state size of $S^1_1$ and $S^1_2$, arriving at the state-reduced CM's, each with 2 states and 11 transitions (see Fig. \ref{Ch-NetworkedPartI-Fig-BasicnetEx1}). To elaborate, using these CM's means: $A_1$ must inform $A_2$ whenever it takes or returns the equipment $E_1$, and $A_2$ reciprocates in turn. Similarly, the CM's $S^2_1$ and $S^2_2$ synthesized using $CMBasicSubnet$ for agents $A_1$ and $A_2$ to cooperatively satisfy $E^2_{\{1,2\}}$ are given in Fig. \ref{Ch-NetworkedPartI-Fig-BasicnetEx2}.

\begin{figure}[ttt]
\centering
\subfigure[CM's of $A_1$ and $A_2$ for $E^1_{\{1,2\}}$]{
    \psfig{file=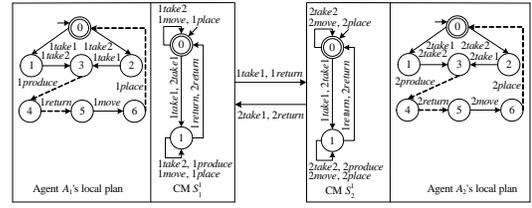,width=7cm} \label{Ch-NetworkedPartI-Fig-BasicnetEx1}
}
\subfigure[CM's of $A_1$ and $A_2$ for $E^2_{\{1,2\}}$]{
    \psfig{file=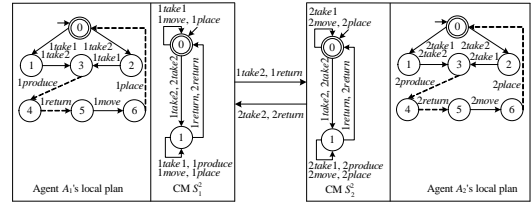,width=7cm} \label{Ch-NetworkedPartI-Fig-BasicnetEx2}
}
\caption{CM's of $A_1$ and $A_2$ for $E^1_{\{1,2\}}$ and $E^2_{\{1,2\}}$.}
\end{figure}
\end{example}

%We refer interested readers to \cite{J-Pham2011} for a more detail discussion and analysis of Procedure $CMBasicSubnet$.

\subsection{Composing Two Basic Subnets} \label{Sect-TwoBasicNets}
We now consider how the CM solutions of two basic subnets can be composed together to obtain a solution for the resultant two-constraint subnet. Given $\mathcal{N}^{\{h,k\}}_2 = \{(J_h, C^h_{J_h}), (J_k, C^k_{J_k})\}$, let $SUP^{\{h,k\}} = Supcon(C^h_{J_h} \parallel C^k_{J_k}, A_{J_h} \parallel A_{J_k})$. We are interested in synthesizing, for each agent $A_i$, a set of CM's $\mathcal{CM}_i$ such that $\underset{i \in J_h \cup J_k}{\parallel} (A_i \parallel CM_i) \equiv SUP^{\{h,k\}}$. Without loss of generality, we assume $J_h \cap J_k \neq \emptyset$.
Otherwise, the two basic subnets contain no common agents and would only need to be synthesized individually.

One simple approach is to reorganize $\mathcal{N}^{\{h,k\}}_2$ into a new subnet consisting of one constraint $C^h_{J_h} \parallel C^k_{J_k}$ for the agent group $\{ A_i \mid i \in J_h \cup J_k \}$. The solution for this reorganized basic subnet can then be obtained by applying $CMBasicSubnet$. This approach, however, has a major drawback: it suffers from exponential complexity of computing the product of all agents $\{ A_i \mid i \in J_h \cup J_k \}$ and constraints $C^h_{J_h}$ and $C^k_{J_k}$. For a large number of agents, this computation may become prohibitively expensive.

Our compositional approach entails designing deconflicting CM's for the agents concerned to resolve any conflict between $\mathcal{N}^h_1$ and $\mathcal{N}^k_1$. The need for additional deconflicting CM's will be clear from the following example.

\begin{example} \label{Ex-Blocking}
For $\mathcal{N}^{\{1,2\}}_2 = \{\mathcal{N}^1_1, \mathcal{N}^2_1\}$, we apply $CMBasicSubnet$ to compute CM's of agents $A_1$ and $A_2$ for $\mathcal{N}^1_1 = (\{1, 2 \}, E^1_{\{1,2\}})$ and $\mathcal{N}^2_1 = (\{1, 2 \}, E^2_{\{1,2\}})$. The CM's $S^1_1$ and $S^1_2$ for $\mathcal{N}^1_1$, and $S^2_1$ and $S^2_2$ for $\mathcal{N}^2_1$, are shown in Figs. \ref{Ch-NetworkedPartI-Fig-BasicnetEx1} and \ref{Ch-NetworkedPartI-Fig-BasicnetEx2}.  However, using only these CM's does not guarantee that $A_1$ and $A_2$ will interact correctly for the subnet $\mathcal{N}^{\{1,2\}}_2$. In fact, the system of $A_1$ and $A_2$ interacting using these CM's contains blocking states. For instance, the event sequence $1take1-2take2$, which is allowed to be executed by the CM's, leads to the blocking situation of each agent holding one equipment and waiting forever to acquire the equipment held by the other agent.
\end{example}

Thus, the local CM's individually constructed for $\mathcal{N}^h_1$ and $\mathcal{N}^k_1$ do not generally constitute a correct solution for $\mathcal{N}^{\{h,k\}}_2$. The reason is that, in general, $SUP^k \parallel SUP^h \not \equiv SUP^{\{h,k\}}$; and whenever this happens, the system of coordinating agents using only their CM's constructed for the individual basic subnets will contain blocking states. We say that $\mathcal{N}^h_1$ and $\mathcal{N}^k_1$ are nonconflicting if $SUP^h \parallel SUP^k$ is nonblocking. Otherwise, they are conflicting. Being nonconflicting means no deconflicting CM's need to be additionally constructed. This motivates the development of a procedure of testing for the nonconflicting of $\mathcal{N}^h_1$ and $\mathcal{N}^k_1$. The simplest way of doing so is to directly compute $SUP^h \parallel SUP^k$ and check whether or not it is a nonblocking automaton. However, this approach is computationally inefficient since it can be shown to have the same complexity order as that of computing the product of all agents and constraints.

Lemma \ref{Ch-NetworkedPartI-Lemma-NonconflictTest2} leads us to a more efficient approach to testing the nonconflict of $\mathcal{N}^h_1$ and $\mathcal{N}^k_1$. This and the next lemmas are formulated in terms of the concepts of language observer and output control consistent (OCC) projection \cite{J-LeiFeng2008}.

\begin{definition} \label{Def-ObserverAndOCC} {\em Observer and OCC Projection \cite{J-LeiFeng2008}:} Given an automaton $A$ with $\Sigma^A = \Sigma^A_{uc} \cup \Sigma^A_{c}$, and $\Sigma \subseteq \Sigma^A$.
\begin{enumerate}
  \item $P_{\Sigma^A,\Sigma}$ is said to be a $L_m(A)$-observer if: $\forall t \in P_{\Sigma^A, \Sigma}(L_m(A)), s \in L(A)$, if $P_{\Sigma^A, \Sigma}(s)$ is a prefix of $t$ then $\exists u \in (\Sigma^A)^*$ such that $su \in L_m(A)$ and $P_{\Sigma^A, \Sigma}(su) = t$.

  \item $P_{\Sigma^A,\Sigma}$ is said to be OCC for $L(A)$ if $\forall s \in L(A)$ of the form $s = s^\prime \sigma_1... \sigma_r$, where $s^\prime$ is either $\varepsilon$ or terminates with an event in $\Sigma$, the following holds: $[\sigma_r \in \Sigma \cap \Sigma^A_{uc}$ and $(\forall k \in I_{r-1}) \sigma_k \in (\Sigma^A - \Sigma)]$ $\Rightarrow [(\forall k \in I_r) \sigma_k \in \Sigma^A_{uc}]$.
\end{enumerate}
\end{definition}

In words, Definition \ref{Def-ObserverAndOCC}.1 asserts that whenever $P_{\Sigma^A,\Sigma}(s)$ can be extended to a string in $P_{\Sigma^A,\Sigma}(L_m(A))$ by catenating to it a string $u^\prime \in \Sigma^*$, the underlying string $s$ can also be extended to a string in $L_m(A)$ by catenating to it a string $u \in (\Sigma^A)^*$ with $P_{\Sigma^A,\Sigma}(u) = u^\prime$. Thus, Definition \ref{Def-ObserverAndOCC}.1 says that every string in the abstract model $P_{\Sigma^A,\Sigma}(A)$ is realizable by the original model $A$.

By Definition \ref{Def-ObserverAndOCC}.2, along every $s \in L(A)$, in between every observable but uncontrollable event that exists and its nearest ``upstream'' observable event  (or otherwise the empty string prior to the ``starting'' event of the string) is a string of uncontrollable and unobservable events. Thus, if $L(A)$ is interpreted as (the behavior of) an underlying system model and $P_{\Sigma^A,\Sigma}(L(A))$ as (the behavior of) the abstracted system model, then, that $P_{\Sigma^A,\Sigma}$ is OCC for $L(A)$ characterizes the semantics that every uncontrollable event in the abstracted model can never be disabled and hence prevented from occurring by disabling controllable events in the underlying model. The abstracted model output $P_{\Sigma^A,\Sigma}(L(A))$ is, in this sense, ``control consistent'' with the underlying model $L(A)$.

\begin{lemma} \label{Ch-NetworkedPartI-Lemma-NonconflictTest2} Let $\Sigma^{\{h,k\}}_{CR} \supseteq \underset{i \in J_k \cap J_h}{\bigcup} \Sigma^{A_i}$ and define $P^h_{CR}$ and $P^k_{CR}$ as projections from $\underset{i \in J_h}{\bigcup} \Sigma^{A_i}$ and $\underset{i \in J_k}{\bigcup} \Sigma^{A_i}$ to $\Sigma^{\{h,k\}}_{CR}$, respectively. Then, if $P^h_{CR}$ is a $L_m(SUP^h)$-observer and $P^k_{CR}$ is a $L_m(SUP^k)$-observer, two basic subnets $\mathcal{N}^h_1$ and $\mathcal{N}^k_1$ are nonconflicting if and only if $P^h_{CR}(SUP^h) \parallel P^k_{CR}(SUP^k)$ is a nonblocking automaton.
\end{lemma}

%\begin{proof}
%See Appendix \ref{Appendix2}.
%\end{proof}

Thus, under the stated sufficiency conditions in Lemma \ref{Ch-NetworkedPartI-Lemma-NonconflictTest2}, testing the nonconflict of$\mathcal{N}^h_1$ and $\mathcal{N}^k_1$ can be reduced to checking whether or not  $P^h_{CR}(SUP^h) \parallel P^k_{CR}(SUP^k)$ is nonblocking. This way, we only need to first compute $P^h_{CR}(SUP^h) \parallel P^k_{CR}(SUP^k)$ instead of $SUP^h \parallel SUP^k$, which results in a computationally cheaper nonconflict test for two reasons. The first is that such automata $P^h_{CR}(SUP^h)$ and $P^k_{CR}(SUP^k)$ can be individually computed in polynomial time \cite{J-KCWong2004}, and the second is that their state sizes are often smaller than those of $SUP^h$ and $SUP^k$, respectively.

Nevertheless, if $\mathcal{N}^h_1$ and $\mathcal{N}^k_1$ are conflicting (due to blocking), we need to design additional deconflicting CM's for the agents concerned to resolve the conflicts between $\mathcal{N}^h_1$ and $\mathcal{N}^k_1$. Together with the local CM's synthesized for $\mathcal{N}^h_1$ and $\mathcal{N}^k_1$, deconflicting CM's will constitute a correct solution for $\mathcal{N}^{\{h,k\}}_2$.  Essentially, deconflicting CM's remove blocking states from $SUP^h \parallel SUP^k$ when used by the agents of subnet $\mathcal{N}^{\{h,k\}}_2$.

In designing deconflicting CM's for coordinating agents, our approach is to first synthesize an automaton as the basis for conflict resolution between two basic subnets, and then ``localize'' it to every agent as the agent's deconflicting CM if the agent shares some events with the conflict resolution (automaton). Formally, an automaton $CR^{\{h,k\}}$ is said to be a conflict resolution for $\mathcal{N}^h_1$ and $\mathcal{N}^k_1$ if $[CR^{\{h,k\}} \parallel SUP^h \parallel SUP^k] \equiv SUP^{\{h,k\}}$.

It can be shown that a conflict resolution for any two basic subnets always exists. Indeed, $CR^{\{h,k\}}$ can be simply computed as $Supcon(G, SUP^h_{J_h} \parallel SUP^k_{J_k})$, where $G$ is a one-state automaton that generates and marks $(\Sigma^{A_{J_h}} \cup \Sigma^{A_{J_k}})^*$. However, similar to the problem of testing the nonconflict of two basic subnets discussed previously, computing $CR^{\{h,k\}}$ as $Supcon(G, SUP^h_{J_h} \parallel SUP^k_{J_k})$  has the same order of complexity as that of $\underset{i \in J_h \cup J_k}{\parallel} A_i$, which is inefficient.

In what follows, we present an efficient approach for computing a conflicting resolution (automaton) for two basic subnets (Lemma \ref{Ch-NetworkedPartI-Thm-LeiFeng}), and using which we propose a conflict resolution algorithm (Procedure $DeconflictBasicSubnet$).

\begin{lemma} \label{Ch-NetworkedPartI-Thm-LeiFeng}
Let $\Sigma^{\{h,k\}}_{CR} \supseteq \underset{i \in J_k \cap J_h}{\bigcup} \Sigma^{A_i}$ and define $P^h_{CR}$ and $P^k_{CR}$ as projections from $\underset{i \in J_h}{\bigcup} \Sigma^{A_i}$ and $\underset{i \in J_k}{\bigcup} \Sigma^{A_i}$ to $\Sigma^{\{h,k\}}_{CR}$, respectively. Then, if $P^h_{CR}$ is a $L_m(SUP^h)$-observer, $P^k_{CR}$ is a $L_m(SUP^k)$-observer, and $\forall i \in J_h \cup J_k$, $P_{\Sigma^{A_i}, \Sigma^{\{h,k\}}_{CR}}$ is OCC for $L(A_i)$, then $CR^{\{h,k\}} = Supcon[G,P^h_{CR}(SUP^h) \parallel P^k_{CR}(SUP^k)]$ is a conflict resolution for $\mathcal{N}^h_1$ and $\mathcal{N}^k_1$, where $G$ is a one-state automaton that generates $(\Sigma^{\{h,k\}}_{CR})^*$ as both the prefix-closed and marked languages.
\end{lemma}

%\begin{proof}
%See Appendix \ref{Appendix3}.
%\end{proof}

Thus, $CR^{\{h,k\}}$ can be computed as $Supcon[G,P^h_{CR}(SUP^h) \parallel P^k_{CR}(SUP^k)]$ if all the conditions stated in Lemma \ref{Ch-NetworkedPartI-Thm-LeiFeng} are satisfied. Importantly, following this approach to compute a conflict resolution, instead of $SUP^h \parallel SUP^k$, we only need to compute the product $P^h_{CR}(SUP^h) \parallel P^k_{CR}(SUP^k)$. Since $P^i_{CR}$, $i \in \{h, k\}$, is a $L_m(SUP^i)$-observer, the state size of $P^i_{CR}(SUP^i)$ is known to be often smaller than that of $SUP^i$.

By Lemma \ref{Ch-NetworkedPartI-Thm-LeiFeng}, a conflict resolution for $\mathcal{N}^h_1$ and $\mathcal{N}^k_1$ can be computed as follows: (i) Initially, let $\Sigma_{J_h \cap J_k} = \underset{i \in J_h \cap J_k}{\bigcup} \Sigma^{A_i}$; (ii) Next, enlarge $\Sigma_{J_h \cap J_k}$ to $\Sigma^{\{h,k\}}_{CR}$ so that all the stated conditions in Lemma \ref{Ch-NetworkedPartI-Thm-LeiFeng} are satisfied; (iii) Then, construct $G$ as a one-state automaton with its only state being both an initial state and a marker state, and with every event in $\Sigma^{\{h, k\}}_{CR}$ self-looped at that state. Thus, $G$ generates $(\Sigma^{\{h, k\}}_{CR})^*$ which is both its prefix-closed and marked languages; (iv) Finally, compute $CR^{\{h,k\}} = Supcon[G, P^h_{CR}(SUP^h) \parallel P^k_{CR}(SUP^k)]$.

Note that the smaller the cardinality of the set $\Sigma^{\{h,k\}}_{CR}$ returned by Step (ii) is, the more economical the computation would be for Step (iv). The problem of finding a minimal cardinality event set $\Sigma^{\{h,k\}}_{CR}$ satisfying every condition in Lemma \ref{Ch-NetworkedPartI-Thm-LeiFeng} has proven to be NP-hard \cite{J-KCWong2004}. However, a polynomial time algorithm exists to synthesize such an event set $\Sigma^{\{h,k\}}_{CR}$ of reasonably small size \cite{J-LeiFeng2008}.

From the foregoing discussion, Procedure $DeconflictBasicSubnet$ is developed to design deconflicting CM's for $\mathcal{N}^h_1$ and $\mathcal{N}^k_1$. It first checks if $\mathcal{N}^h_1$ and $\mathcal{N}^k_1$ are nonconflicting by applying Lemma \ref{Ch-NetworkedPartI-Lemma-NonconflictTest2} (Step 1). If they are, then no deconflicting CM is needed. Otherwise, Lemma \ref{Ch-NetworkedPartI-Thm-LeiFeng} is applied to compute a conflict resolution $CR^{\{h,k\}}$ for the two subnets (Step 2). Next, in Step 3, the procedure determines whether or not an agent $A_i$ needs to take part in resolving the conflict between the subnets, i.e., if $\Sigma^{CR^{\{h,k\}}} \cap \Sigma^{A_i} \neq \emptyset$. If so, it computes for $A_i$ a deconflicting CM $S^{\{h,k\}}_i$. Note that such a deconflicting CM could simply be taken as $CR^{\{h,k\}}$. However, to achieve economy of implementation, it uses $CMreduce$ to obtain a reduced CM $S^{\{h,k\}}_i = CMreduce(CR^{\{h,k\}}, A_i)$. In the worst case, $\Sigma^{\{h,k\}}_{CR} = \underset{i \in J_h \cup J_k}{\bigcup} \Sigma^{A_i}$ and $DeconflictBasicSubnet$ has to compute the synchronous product of all agents and constraints in the two subnets. It therefore has exponential complexity. However, $DeconflictBasicSubnet$ is often efficient in practice since $\Sigma^{\{h,k\}}_{CR}$ is often a strict subset of $\underset{i \in J_h \cup J_k}{\bigcup} \Sigma^{A_i}$.

\DontPrintSemicolon
\begin{procedure}
{\small
\textbf{Output:} A deconflicting CM $S^{\{h,k\}}_i$ for agent $A_i$ to resolve the conflict between $\mathcal{N}^h_1$ and $\mathcal{N}^k_1$

\Begin{
\textbf{Step 1}: Check if $\mathcal{N}^h_1$ and $\mathcal{N}^k_1$ are nonconflicting:
    \begin{itemize}
        \item \textbf{Step 1a} Let $\Sigma^{\{h,k\}}_{CR} = \underset{i \in J_h \cap J_k}{\bigcup} \Sigma^{A_i}$. Enlarge $\Sigma^{\{h,k\}}_{CR}$ so \\ that $P^h_{CR}$ and $P^k_{CR}$ become a $L_m(SUP^h)$-observer and $L_m(SUP^k)$-observer, respectively ($P^h_{CR}$ and $P^k_{CR}$ are projections from  $\underset{i \in J_h}{\bigcup} \Sigma^{A_i}$ and $\underset{i \in J_k}{\bigcup} \Sigma^{A_i}$ to $\Sigma^{\{h,k\}}_{CR}$);
        \item \textbf{Step 1b} If $\overline{L_m(P^h_{CR}(SUP^h)) \parallel L_m(P^k_{CR}(SUP^k))}$ $= L(P^h_{CR}(SUP^h)) \parallel L(P^k_{CR}(SUP^k))$, i.e., $\mathcal{N}^h_1$
            and \\ $\mathcal{N}^k_1$ are nonconflicting, no deconflicting CM is needed. \\ Otherwise, go to Step 2 to
            design deconflicting CM's;
    \end{itemize}
\textbf{Step 2}: Compute $CR^{\{h,k\}}$\;
    \begin{itemize}
        \item \textbf{Step 2a} Enlarge $\Sigma^{\{h,k\}}_{CR}$ so that $P^h_{CR}$ is a $L_m(SUP^h)$-observer, $P^k_{CR}$  is a
             $L_m(SUP^k)$-observer,\\ and $\forall i \in J_h \cup J_k$, $P_{\Sigma^{A_i}, \Sigma^{\{h,k\}}_{CR}}$ is OCC for $L(A_i)$;

        \item \textbf{Step 2b} Construct $G$ as a one state automaton with \\ its only state being both an initial state and a marker \\ state, and with every event in $\Sigma^{\{h, k\}}_{CR}$ self-looped at that \\ state;

        \item \textbf{Step 2c} Compute $CR^{\{h,k\}} = Supcon[G, P^h_{CR}(SUP^h) \parallel P^k_{CR}(SUP^k)]$;
    \end{itemize}
\textbf{Step 3}: For each agent $A_i$ in the subnet $\mathcal{N}^{\{h,k\}}_2$, if $\Sigma^{CR^{\{h,k\}}} \cap \Sigma^{A_i} \neq \emptyset$, compute for $A_i$ a deconflicting CM $S^{\{h,k\}}_i = CMreduce(CR^{\{h,k\}}, A_i)$; \;
}
\caption{$DeconflictBasicSubnet$($\mathcal{N}^h_1$, $\mathcal{N}^k_1$) \label{Ch-NetworkedPartI-Procedure-MCSA2}}
}
\end{procedure}

\begin{lemma} \label{Ch-NetworkedPartI-Lemma-DeconflictBasicSubnet}
For $i \in J_h \cup J_k$, let $S^{\{h,k\}}_i$ be the deconflicting CM computed for agent $A_i$ in Step 3 of $DeconflictBasicSubnet$, or trivially a one-state automaton that generates and marks $(\Sigma^{A_i})^*$ if no deconflicting CM is needed for $A_i$, either because $\mathcal{N}^h_1$ and $\mathcal{N}^k_1$ are nonconflicting or because $\Sigma^{CR^{\{h,k\}}} \cap \Sigma^{A_i} = \emptyset$. Then, $\underset{i \in J_h \cup J_k}{\parallel} (A_i \parallel S^{\{h,k\}}_i) \equiv CR^{\{h,k\}}$.
\end{lemma}

%\begin{proof}
%See Appendix \ref{Appendix4}.
%\end{proof}

Theorem \ref{Ch-NetworkedPartI-Thm-Subnet} formally summarizes how we can compose (the solution CM's of) two basic subnets $\mathcal{N}^h_1$ and $\mathcal{N}^k_1$ to form (a CM solution set for) the two-constraint subnet $\mathcal{N}^{\{h,k\}}_2$.

\begin{theorem} \label{Ch-NetworkedPartI-Thm-Subnet}
For $i \in J_h \cup J_k$, let $\mathcal{CM}_i$ be the CM set for agent $A_i$ computed as follows: (i) $\mathcal{CM}_i$ includes every CM computed for $A_i$ when applying $CMBasicSubnet$ for $\mathcal{N}^h_1$ and $\mathcal{N}^k_1$, and (ii) $\mathcal{CM}_i$ includes every deconflicting CM computed for $A_i$ when applying $DeconflictBasicSubnet$ to resolve the conflict that exists between $\mathcal{N}^h_1$ and $\mathcal{N}^k_1$. Then $\underset{i \in J_h \cup J_k}{\parallel} (A_i \parallel CM_i) \equiv SUP^{\{h,k\}}$, where $CM_i$ is a synchronous product of all CM's in $\mathcal{CM}_i$.
\end{theorem}

\subsection{Composing Two Arbitrary Subnets} \label{Sect-TwoArbitraryNets}
With slight modifications, the theoretical results presented in the previous section can be generalized to composing two subnets $\mathcal{N}^{\mathcal{S}_x}_{x}$ and $\mathcal{N}^{\mathcal{S}_y}_{y}$ of sizes $x, y \in I_m$, to form a larger $(x + y)$-constraint subnet. In doing so, we follow the same composition logic, i.e., we first synthesize the CM's for each individual subnet, and then design, if necessary, deconflicting CM's for the agents to resolve the conflict between the two subnets. A procedure called $DeconflictSubnet$ for $\mathcal{N}^{\mathcal{S}_x}_{x}$ and $\mathcal{N}^{\mathcal{S}_y}_{y}$ is developed. It is almost identical to but extends $DeconflictBasicSubnet$ based on a straightforward generalization of Lemma \ref{Ch-NetworkedPartI-Thm-LeiFeng}.

\section{Multiagent Conflict Resolution Planning} \label{Sect-ConflictResolutionPlanning}
This section fills in Step 2.1 of our compositional synthesis approach, presenting the formalism and algorithms for the representation and generation conflict resolution plans.

\subsection{AND/OR Graph for Conflict Resolution Plans} \label{Ch-NetworkedPartII-Sect-Representation}
\begin{definition} \label{Ch-NetworkedPartII-Def-SubnetDecomposition}
Given a DCSN $\mathcal{N}$ consisting of $m$ basic subnets $\mathcal{N}^1_1$, ... $\mathcal{N}^m_1$, a subnet-decomposition $\Phi$ is a set of subnets of $\mathcal{N}$ such that: 1) Every element subnet of $\Phi$ is constraint-connected, 2) every basic subnet of $\mathcal{N}$ is contained in one of the elements of $\Phi$, and 3) there is no basic subnet of $\mathcal{N}$ that is contained in two different elements of $\Phi$.
\end{definition}
%}
It follows that a conflict resolution plan for $\mathcal{N}$ is a sequence of transitions of subnet-decompositions, starting with $\Phi_I = \{\mathcal{N}^1_1,\mathcal{N}^2_1, ... ,  \mathcal{N}^m_1\}$ and ending with $\Phi_F = \{ \mathcal{N} \}$. $\Phi_I$ characterizes the situation in which all the basic subnets are ``disconnected'' from each other, and $\Phi_F$ characterizes the situation where all the basic subnets are already deconflicted together to form the complete DCSN $\mathcal{N}$. Each transition from one subnet-decomposition to another characterizes an operation of deconflicting (the CM solutions of) subnets to form (a CM solution of) a larger subnet. A conflict resolution plan should only include transitions that correspond to resolving conflicts of subnets that contain common agents, since subnets that contain no common agents are trivially nonconflicting. Generating a conflict resolution plan for $\mathcal{N}$ is then equivalent to searching for a path of subnet-decomposition transitions from $\Phi_I$ to $\Phi_F$.

Observe that a conflict resolution planning sequence for a DCSN $\mathcal{N}$ is a reversal of a successive decomposition, starting with $\mathcal{N}$, of constraint-connected component subnets until only basic subnets remain. This suggests that the forward search problem of generating conflict resolution plans for a DCSN $\mathcal{N}$ can be addressed as a backward search problem of successively decomposing $\mathcal{N}$ into pairs of constraint-connected component subnets until only basic subnets are left. The space of all possible conflict resolution plans for $\mathcal{N}$ can therefore be generated by enumerating all possible ways of successively decomposing $\mathcal{N}$ this way.

Because there are many subnet-decompositions that can be made from the same DCSN, the branching factor from the initial state $\Phi_I$ to the goal state $\Phi_F$ is greater than that from $\Phi_F$ to $\Phi_I$. A backward search is, therefore, often more efficient than a forward search for the conflict resolution planning problem.

AND/OR graphs \cite{B-StuartRussell2003} are suitable in representing decomposable problems. By recognizing that conflict resolution plans for a DCSN can be generated by enumerating all possible ways of successively decomposing it, Definition \ref{Ch-NetworkedPartII-Def-ANDORGraph} proposes a representation using AND/OR graphs for the conflict resolution plans of a DCSN.

\begin{definition} \label{Ch-NetworkedPartII-Def-ANDORGraph}
The AND/OR graph of conflict resolution plans for a DCSN $\mathcal{N}$ is a hyper-graph $T_\mathcal{N} = (S_\mathcal{N}, H_\mathcal{N})$, where
\begin{enumerate}
  \item $S_\mathcal{N}$ is the set of nodes of $T_\mathcal{N}$ and defined as $S_\mathcal{N} = \{ \mathcal{N}_r \subseteq \mathcal{N} \mid  \mathcal{N}_r \mbox{ is constraint-connected }\}$.
  \item $H_\mathcal{N}$ is the set of hyper-edges of $T_\mathcal{N}$ and defined as $H_\mathcal{N} = \{ (\mathcal{N}_{r_1}, (\mathcal{N}_{r_2}, \mathcal{N}_{r_3})) \in S_\mathcal{N} \times (S_\mathcal{N} \times S_\mathcal{N}) \mid \mathcal{N}_{r_2} \cap \mathcal{N}_{r_3} \neq \emptyset \mbox{ and } \mathcal{N}_{r_1} = \mathcal{N}_{r_2} \cup \mathcal{N}_{r_3} \}$.%\hspace{\fill}$\blacksquare$
\end{enumerate}
\end{definition}

The nodes in the AND/OR graph $T_\mathcal{N}$ represent constraint-connected  subnets of $\mathcal{N}$, and each of the hyper-edges is a pair $(\mathcal{N}_{r_1}, (\mathcal{N}_{r_2}, \mathcal{N}_{r_3}))$ denoting the decomposition of subnet $\mathcal{N}_{r_1}$ into two component subnets $\mathcal{N}_{r_2}$ and $\mathcal{N}_{r_3}$, or equivalently, the composition of $\mathcal{N}_{r_2}$ and $\mathcal{N}_{r_3}$ into $\mathcal{N}_{r_1}$. A hyper-edge points from a node representing a subnet to two nodes representing the component subnets. The node that represents the complete DCSN $\mathcal{N}$ is referred to as the root node and denoted by $n_{root}$, and the nodes representing basic subnets of $\mathcal{N}$ are referred to as the leaf nodes. The set of all leaf nodes of $T_\mathcal{N}$ is $\{ \mathcal{N}_1 \subseteq \mathcal{N} \mid \mathcal{N}_1 \mbox{ is a basic subnet of } \mathcal{N} \}$, and is denoted by $\Theta_{leaf}$.

In what follows, a conflict resolution plan for $\mathcal{N}$ is represented by a tree in $T_\mathcal{N}$ that starts at $n_{root}$ and terminates at $\Theta_{leaf}$. Formally, a tree $tree$ in the AND/OR graph $T_\mathcal{N} = (S_\mathcal{N}, H_\mathcal{N})$, starting at a node $n_I \in S_\mathcal{N}$ and terminating at a set of nodes $\Theta \subseteq S_\mathcal{N}$, can be described recursively as follows.

\begin{itemize}
  \item If $n_I \in \Theta$, $tree$ contains only one node $n_I$ and no edge, and we write $tree = (n_I)$.
  \item Otherwise, $tree$ contains the node $n_I$, an edge $h = (n_I, (n_1, n_2)) \in H_\mathcal{N}$, and the nodes and edges of two trees $tree_1$ and $tree_2$. Each tree $tree_i$, $i \in \{1,2\}$, starts from one of $n_I$'s two successors, $n_i$, and terminates at some $\Theta_i \subseteq \Theta$, where  $\Theta_1$ and $\Theta_2$ are disjoint and $\Theta_1 \cup \Theta_2 = \Theta$. In this case, we write $tree = (n_I, h, tree_1, tree_2)$.
\end{itemize}

The set of all trees starting from $n_I$ and terminating at $\Theta$ is denoted by $Trees(n_I, \Theta)$. If $tree \in Trees(n_I, \Theta)$, $n_I$ is called the root node of $tree$ and a node in $\Theta$ called a terminal node of $tree$. Whenever the set of terminal nodes is arbitrary, the set of trees starting from a node $n_I$ is simply denoted by $Trees(n_I, -)$, and the set of all trees of $T_\mathcal{N}$ is denoted by $Trees(-,-)$.

A tree in $Trees(n_{root}, \Theta_{leaf})$ is said to be complete. Formally then, a complete tree is a conflict resolution plan. Any tree in $T_\mathcal{N}$ whose root node is not $n_{root}$ or whose leaf nodes are not all in $\Theta_{leaf}$ is called a non-complete tree. A non-complete tree is a subgraph %\footnote{A graph $G = (V, E)$ is said to be a subgraph of another graph $G^\prime = (V^\prime,E^\prime)$ if $V \subseteq V^\prime$ and $E \subseteq E^\prime$, namely, the nodes and edges of $G$ are all contained in $G^\prime$.}
of one or more complete trees. A non-complete tree whose root node is $n_{root}$ is called a partial tree. In what follows, a tree in $Trees(n_{root},-)$ is a partial conflict resolution plan.

\subsection{AND/OR Graph Generation of Conflict Resolution Plans} \label{Ch-NetworkedPartII-Sect-Generation}
We now present an algorithm for generating the AND/OR graph representation of conflict resolution plans. Our algorithm takes as input a DCSN and generates as output the AND/OR graph representation of conflict resolution plans for the DCSN.

The basic idea of our algorithm is to first enumerate all possible decompositions of a DCSN $\mathcal{N}$ into two constraint-connected  component subnets. Each such decomposition corresponds to an edge of the AND/OR graph $T_\mathcal{N}$ connecting the root node representing $\mathcal{N}$ to two nodes, with each representing a component subnet. The same decomposition process is then repeated for each of the component subnets, which are component DCSN's, until only basic subnets are left. Recursive decomposition lends itself to straightforward AND/OR graph construction of all conflict resolution plans.

To facilitate the systematic enumeration of all possible decompositions of a subnet in a DCSN, we first convert the DCSN to a constraint relational network (CRN). In essence, the CRN of a DCSN, formally defined in Definition \ref{Ch-NetworkedProblemII-Def-CRN} below, is a constraint relational model which explicitly relates every pair of inter-agent constraints whose induced agent groups overlap.

\begin{definition} \label{Ch-NetworkedProblemII-Def-CRN}
The constraint relational network (CRN) $\mathcal{CRN}_r$ of a $r$-constraint subnet $\mathcal{N}^{\mathcal{S}_r}_r = \{(J_k, C^k_{J_k}) \mid k \in \mathcal{S}_r \}$ is a tuple $(\mathcal{C}_r, \mathcal{R}_r)$, where  $\mathcal{C}_r = \{ C^k_{J_k} \mid k \in \mathcal{S}_r \}$ is the constraint set of size $r$ in $\mathcal{N}_r$ and $\mathcal{R}_r \subseteq \mathcal{C}_r \times \mathcal{C}_r$ is a relation over $\mathcal{C}_r$, such that $(\forall C^k_{J_k}, C^h_{J_h} \in \mathcal{C}_r)[(C^k_{J_k}, C^h_{J_h}) \in \mathcal{R}_r \Leftrightarrow (J_k \cap J_h \neq \emptyset)]$.%\hspace{\fill}$\blacksquare$
\end{definition}

By Definition \ref{Ch-NetworkedProblemII-Def-CRN}, two constraints $C^k_{J_k}$ and $C^h_{J_h}$ are related if their induced agent groups are overlapping, i.e., $J_k \cap J_h \neq \emptyset$, meaning that there is at least one agent $A_i$, where $i \in J_k \cap J_h$, that belongs to both the basic subnets $(J_k, C^k_{J_k})$ and $(J_h, C^h_{J_h})$. In other words, $A_i$ has to coordinate on $C^k_{J_k}$ with some agents, and on $C^h_{J_h}$ with some other agents. As already discussed in Section \ref{Sect-TwoBasicNets}, conflicts between (the agents in) such a pair of subnets may arise, and hence, there is a need to check for and resolve any conflict when composing the subnets.

Graphically, a CRN can be represented by an undirected graph with constraints represented by nodes, and the relation between two agent-related constraints $C^k_{J_k}$ and $C^h_{J_h}$ by an edge that connects the corresponding two nodes and is labeled with the agent group overlap between the subnets $(J_k, C^k_{J_k})$ and $(J_h, C^h_{J_h})$.

Observe that enumerating all possible decompositions of a subnet $\mathcal{N}_r$ into two constraint-connected  subnets can be done by enumerating all possible cut-sets\footnote{In a connected graph $G=(V,E)$, a cut-set \cite{B-NarsinghDeo1974} is a set of edges $E^\prime \subseteq E$ such that the removal of $E^\prime$ from $G$ disconnects $G$ and the removal of any strict subset of $E^\prime$ does not disconnect $G$. Since a cut-set $E^\prime$ always ``cuts'' $G$ into two parts, it may be conveniently represented as $(V_1, V_2)$, where $V_1$ and $V_2$ are the sets of vertices belonging to these two parts. Let $T$ be a spanning tree of $G$. Then a ``fundamental'' cut-set of $G$ is defined as a cut-set that contains exactly one branch of $T$.  Defining the ring sum operation $\oplus$ of two arbitrary sets $A$ and $B$ as $A \oplus B = (A \cup B) - (A \cap B)$, it has been shown that any cut-set of $G$ has the form $E_1 \oplus E_2 \oplus ... \oplus E_z$ that is not a union of edge-disjoint cut-sets, where $z \geq 2$ is arbitrary and $E_1, ...,E_z$ are different fundamental cut-sets of $G$. Thus, a formal approach to generate all cut-sets of $G$ is to (i) construct a spanning tree, (ii) generate the set of fundamental cut-sets for the spanning tree, and then (iii) properly combine these fundamental cut-sets to get a new cut-set.}
of its CRN $\mathcal{CRN}_r$. Specifically, consider a cut-set $(\mathcal{C}_x, \mathcal{C}_y)$ that decomposes $\mathcal{CRN}_r$ into two parts, where $\mathcal{C}_x$ and $\mathcal{C}_y$ are the two disjoint sets of vertices of $\mathcal{CRN}_r$ belonging to these two parts. Write $\mathcal{N}_x \sim  \mathcal{C}_x$ and $\mathcal{N}_y \sim  \mathcal{C}_y$ to denote respectively that $\mathcal{N}_x$ and $\mathcal{N}_y$ are the component subnets induced by $\mathcal{C}_x$ and $\mathcal{C}_y$, namely $\mathcal{N}_x =  \{ (J_k, C^k_{J_k}) \mid C^k_{J_k} \in \mathcal{C}_x \}$ and $\mathcal{N}_y =  \{ (J_k, C^k_{J_k}) \mid C^k_{J_k} \in \mathcal{C}_y \}$. Then $\mathcal{N}_x$ and $\mathcal{N}_y$ are two constraint-connected  component subnets decomposed from $\mathcal{N}_r$. Conversely, any decomposition of $\mathcal{N}_r$ into two constraint-connected  component subnets $\mathcal{N}_x$ and $\mathcal{N}_y$ corresponds to a cut-set $(\mathcal{C}_x, \mathcal{C}_y)$ of $\mathcal{CRN}_r$, with $\mathcal{N}_x \sim  \mathcal{C}_x$ and $\mathcal{N}_y \sim  \mathcal{C}_y$.

From the foregoing observation, Procedure $GenerateANDORGraph$ details the steps to generate an AND/OR graph representation of conflict resolution plans for a given DCSN $\mathcal{N}$. If $\mathcal{N}$ is a basic subnet, the procedure simply returns an empty AND/OR graph (Step 1), otherwise it converts $\mathcal{N}$ to the a CRN $\mathcal{CRN}$, and computes $CutSets$ as the set of all cut-sets of $\mathcal{CRN}$ (Step 2). In Step 3, the procedure uses the cut-sets to recursively construct the AND/OR graph representation of conflict resolution plans.

\DontPrintSemicolon
\begin{procedure}
{\small
\textbf{Output:} An AND/OR graph $T_\mathcal{N} = (S_\mathcal{N}, H_\mathcal{N})$ of conflict resolution plans for $\mathcal{N}$, initialized with $S_\mathcal{N} = \emptyset$ and $H_\mathcal{N} = \emptyset$

\Begin{
    \textbf{Step 1}: If $\mathcal{N}$ contains only one basic subnet then return; otherwise, convert $\mathcal{N}$ into a $\mathcal{CRN} = (\mathcal{C}, \mathcal{R})$; \;

    \textbf{Step 2}: Compute $CutSets$ as the set of all cut-sets of $\mathcal{CRN}$; \;

    \textbf{Step 3} \While{$CutSets \neq \emptyset$}{
        \textbf{Step 3a} Remove a cut-set $(\mathcal{C}_x, \mathcal{C}_y)$ from $CutSets$. Let $\mathcal{N}_x \sim \mathcal{C}_x$ and $\mathcal{N}_y \sim \mathcal{C}_y$; \;
        \textbf{Step 3b} Add nodes and an edge to $T$: $S_\mathcal{N} = S_\mathcal{N} \cup \{ \mathcal{N}_x, \mathcal{N}_y, \mathcal{N}_x \cup \mathcal{N}_y \}$, $H_\mathcal{N} \cup \{ (\mathcal{N}_x \cup \mathcal{N}_y, \mathcal{N}_x, \mathcal{N}_y) \}$;\;
        \textbf{Step 3c} For $r \in \{x, y \}$, $GenerateANDORGraph(\mathcal{N}_r)$; \;
    }
}
\caption{$GenerateANDORGraph$($\mathcal{N}$)}
}
\end{procedure}

Based on the foregoing discussion, $GenerateANDORGraph$ is correct and complete in the sense that it correctly generates, for a DCSN $\mathcal{N}$, an AND/OR graph that completely encompasses all possible conflict resolution plans for $\mathcal{N}$.

The amount of computation involved depends on the number of basic subnets of the input DCSN and its connectivity structure, which both affect the number of cut-sets of the CRN of $\mathcal{N}$ and that of the CRN of each successively decomposed subnet. A complexity evaluation of the algorithm has been conducted, which shows that in general, the more basic subnets and the more ``connected'' they are in an input DCSN, the higher the amount of computation incurred. Presented elsewhere \cite{PhD-Pham2011}, a complexity evaluation of the algorithm has been conducted, which shows that in general, the more basic subnets and the more ``connected'' they are in an input DCSN, the higher the amount of computation incurred. Given a DCSN with $m$ basic subnets, the worst-case complexity of $GenerateANDORGraph ranges$ from $O(m^2)$ to $O(2^m)$.

In practice, based on some criterion, the cut-sets may be subjected to some acceptance tests in Step 3a, and only accepted cut-sets are passed on to Steps 3b and 3c. Such tests can be developed to generate conflict resolution plans which must also satisfy some problem-dependent conditions. For example, a particular multiagent coordination system may contain some subnets which need to be able to run standalone from time to time. To support this standalone operation, we need to guarantee multiagent nonblocking reconfigurability for every standalone subnet; in other words, at the outset, we need to guarantee that agents in a standalone subnet can always maintain nonblockingness of their subnet's coordination tasks during runtime, after a system network reconfiguration of simply unloading all other agent and CM models not relevant to the subnet. This has significant implications in generating and selecting conflict resolution plans. Given a DCSN containing standalone subnets, not all of which are basic, we would need an AND/OR graph plan representation that must include only decompositions in which each of these subnets is wholly contained in a child node of the graph, whenever it is part of a bigger subnet in the parent node. Executing such plans forward can then guarantee multiagent nonblocking reconfigurability. A simple cut-set acceptance test can be developed to generate such AND/OR graph plans.

\subsection{Selection of An Optimal Conflict Resolution Plan} \label{Ch-NetworkedPartII-Sect-Selection}
\subsubsection{General Heuristic Search for An Optimal Conflict Resolution Plan} \label{Ch-NetworkedPartII-Sect-HeuristicSearch}
To select an optimal conflict resolution plan for a given DCSN $\mathcal{N}$, in addition to the ability to traverse the space of all possible conflict resolution plans provided by $T_\mathcal{N}$, there is a need for an optimization metric to access, or rank, the quality of individual plans.

Since a conflict resolution plan is a tree in $T_\mathcal{N}$ that starts from $n_{root} \in S_\mathcal{N}$ and terminates at $\Theta_{leaf} \subseteq S_\mathcal{N}$, an optimization metric for plan selection is simply a real function $F: Trees(n_{root}, \Theta_{leaf}) \rightarrow \mathbb{R}$, where $\mathbb{R}$ is the set of real numbers. We assume a minimization problem, and interpret a better conflict resolution plan as a plan with lower $F$-value. Thus if $tree_1$, $tree_2 \in Trees(n_{root}, \Theta_{leaf})$ with $F(tree_1) <  F(tree_2)$, then the conflict resolution plan $tree_1$ is preferable to $tree_2$.

Selecting an optimal plan can be made algorithmically using a heuristic defined on the set of partial trees $Trees(n_{root}, \Theta_{leaf})$ in $T_\mathcal{N}$ for a given optimization metric.

\begin{definition} \label{Ch-NetworkedPartII-Def-Heuristic}
A heuristic (for an optimization metric $F: Trees(n_{root}, \Theta_{leaf}) \rightarrow \mathbb{R}$) is a real function $H: Trees(n_{root},-) \rightarrow \mathbb{R}$ such that $(\forall tree \in Trees(n_{root}, \Theta_{leaf})) H(tree) = F(tree)$.%\hspace{\fill}$\blacksquare$
\end{definition}

Given a partial tree $ptree \in Trees(n_{root},-)$, the heuristic value $H(ptree)$ shall be used in our algorithm as an estimation of the $F$-value of the best conflict resolution plan $tree \in Trees(n_{root}, \Theta_{leaf})$ that encompasses the partial plan $ptree$. Heuristic $H$ is said to be admissible if the $H$-value of an arbitrary partial tree always underestimates the $F$-value of any complete tree encompassing it, as formalized in Definition \ref{Ch-NetworkedPartII-Def-Admissible}.

\begin{definition} \label{Ch-NetworkedPartII-Def-Admissible}
A heuristic $H: Trees(n_{root}, -) \rightarrow \mathbb{R}$ is said to be admissible (for an optimization  metric $F: Trees(n_{root}, \Theta_{leaf}) \rightarrow \mathbb{R}$) if,  for an arbitrary partial $ptree \in Trees(n_{root}, -)$ and every complete tree $tree \in Trees(n_{root}, \Theta_{leaf})$ for which $ptree$ is a subgraph of, $H(ptree) \leq F(tree)$.%\hspace{\fill}$\blacksquare$
\end{definition}

We can now formally present our plan selection algorithm. Given an admissible heuristic $H$ for some optimization metric $F$, Procedure $HeuristicPlanSelection$ details the steps to select an optimal conflict resolution plan for a DCSN $\mathcal{N}$ from the AND/OR graph $T_\mathcal{N}$. The procedure returns a complete tree of $T_\mathcal{N}$ with the lowest $F$-value, and is thus an optimal conflict resolution plan for $\mathcal{N}$.

\DontPrintSemicolon
\begin{procedure}
{\small
\textbf{Input:} AND/OR graph of conflict resolution plans $T_\mathcal{N} = (S_\mathcal{N}, H_\mathcal{N})$ for DCSN $\mathcal{N}$ and an admissible heuristic $H: Trees(n_{root}, -) \rightarrow \mathbb{R}$

\textbf{Output:} A tree in $Trees(n_{root}, \Theta_{leaf})$ with the lowest $F$-value, which is an optimal conflict resolution plan for $\mathcal{N}$

\Begin{
    \textbf{Step 1}: Create a partial tree $ptree$ which contains only the root node $n_{root}$; \;
    \textbf{Step 2}: Compute the heuristic value $H(ptree)$ and put $ptree$ into a queue $Q$; \;
    \textbf{Step 3}: \While{$Q \neq \emptyset$}{
        \textbf{Step 3a} Extract from $Q$ a tree with the lowest $H$-value and call it $ptree$; \;
        \textbf{Step 3b} If $ptree \in Trees(n_{root}, \Theta_{leaf})$, return it as a solution; \;
        \textbf{Step 3c} Otherwise, select a terminal node $n$ of $ptree$ that is not in $\Theta_{leaf}$; \;
        \textbf{Step 3d} \For{each edge $(n, (n_1, n_2)) \in H_\mathcal{N}$}{
            \textbf{Step 3d1} Create a new partial tree $ntree$ whose nodes are those of $ptree$ plus $n_1$ and $n_2$, and whose edges are those of $ptree$ plus $(n, (n_1, n_2))$; \;
            \textbf{Step 3d2} Compute $H(ntree)$ and put $ntree$ into $Q$; \;
        }
    }
}
\caption{$HeuristicPlanSelection$($T_\mathcal{N}, H$)}
}
\end{procedure}

$HeuristicPlanSelection$ maintains a priority queue $Q$ that contains partial trees of $T_\mathcal{N}$, ranked by their heuristic $H$-value. In Steps 1 and 2, a partial tree that contains only the root node $n_{root}$ is created and put into $Q$. Each time through the while loop of Step 3, a tree with the lowest $H$-value is extracted from $Q$ (Step 3a), and is returned as a solution if it is a complete tree (Step 3b), or otherwise expanded (Steps 3c and 3d). The expanded trees are then put into $Q$ for further examination (Step 3d2). %The correctness of $HeuristicPlanSelection$ is formally stated in Theorem \ref{Ch-NetworkedPartII-Prop-Heuristic}.

\begin{theorem} \label{Ch-NetworkedPartII-Prop-Heuristic}
If $H$ is an admissible heuristic for $F$, then $HeuristicPlanSelection$ returns a complete tree in $Trees(n_{root}, \Theta_{leaf})$ with the lowest $F$-value.
\end{theorem}
%\begin{proof}
%See Appendix \ref{Appendix6}
%\end{proof}

\subsubsection{Reducing Execution Time through Parallel Compositions of Subnets} \label{Ch-NetworkedPartII-Sect-ParallelHeuristic}
We now introduce a criterion to evaluate and select conflict resolution plans. The criterion is to maximize the simultaneous execution of operations for subnet composition. An optimization metric to rank the plans quantitatively based on this criterion is formulated, and an admissible heuristic of this metric is designed for $HeuristicPlanSelection$. Importantly, the selected plan provides the opportunity to maximize the parallel use of available computing resources in simultaneous subnet compositions, and can often be executed in minimal total execution time.

Over a conflict resolution planning tree in the AND/OR graph $T_\mathcal{N}$, the measure of simultaneity of execution supported in the operations of subnet composition can be quantified by the depth of the tree, defined recursively as follows. $(\forall tree \in Trees(-,-))Depth(tree)=0$ if $tree = (n_I)$, and $Depth(tree)= 1 + max(Depth(tree_1), Depth(tree_2))$ if $tree = (n_I, h, tree_1, tree_2)$.

Using this measure, the optimization metric is defined as: $F_p: Trees(n_{root}, \Theta_{leaf}) \rightarrow \mathbb{N}$ such that $F(tree) = Depth(tree)$, where $\mathbb{N} = \{0, 1, 2,... \}$ is the set of natural numbers.

We now design an admissible heuristic $H_p$ for $F_p$. Recall that the set of nodes of $T_\mathcal{N}$ is $S_\mathcal{N} = \{ \mathcal{N}_r \subseteq \mathcal{N} \mid  \mathcal{N}_r \mbox{ is constraint-connected }\}$, namely, each node of $T_\mathcal{N}$ represents a constraint-connected  subnet of $\mathcal{N}$. For each $n \in S_\mathcal{N}$, let $NumBasicSubnet(n)$ denote the number of basic subnets in the constraint-connected  subnet represented by node $n$.

Let $H^\prime_p$ be a real function on $Trees(-,-)$, defined recursively as follows: $H^\prime_p(tree)= \log_2(NumBasicSubnet(n_{I}))$ if $tree = (n_I)$, and $H^\prime_p(tree)=1 + max(H^\prime_p(tree_1), H^\prime_p(tree_2))$ if $tree = (n_I, h, tree_1, tree_2)$. Then an admissible heuristic $H_p$ for $F_p$ can be specified as $H_p: Trees(n_{root}, -) \rightarrow \mathbb{R}$ such that $H_p(ptree) = H^\prime_p(ptree)$.

%The proof of Lemma \ref{Ch-NetworkedPartII-Sect-LemmaAdmissible} unravels the intuition behind the design of $H_p$.

\begin{lemma} \label{Ch-NetworkedPartII-Sect-LemmaAdmissible}
$H_p$ is an admissible heuristic for $F_p$.
\end{lemma}
%\begin{proof}
%See Appendix \ref{Appendix7}.
%\end{proof}

Thus, by Lemma \ref{Ch-NetworkedPartII-Sect-LemmaAdmissible}, Heuristic $H_p$ can be incorporated into $HeuristicPlanSelection$ for the selection of a plan with the lowest $F_p$ value.

\begin{figure}[!ht]
\centering
\subfigure[CRN]
{
    \psfig{file=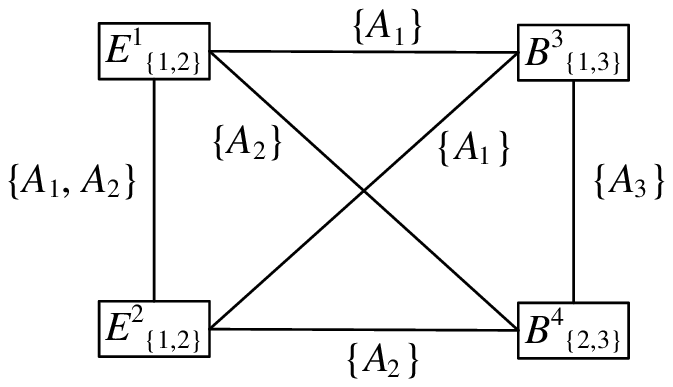,width=4cm} \label{Ch-NetworkedPartII-Fig-CRN}
}
\subfigure[Plan 1]
{
    \psfig{file=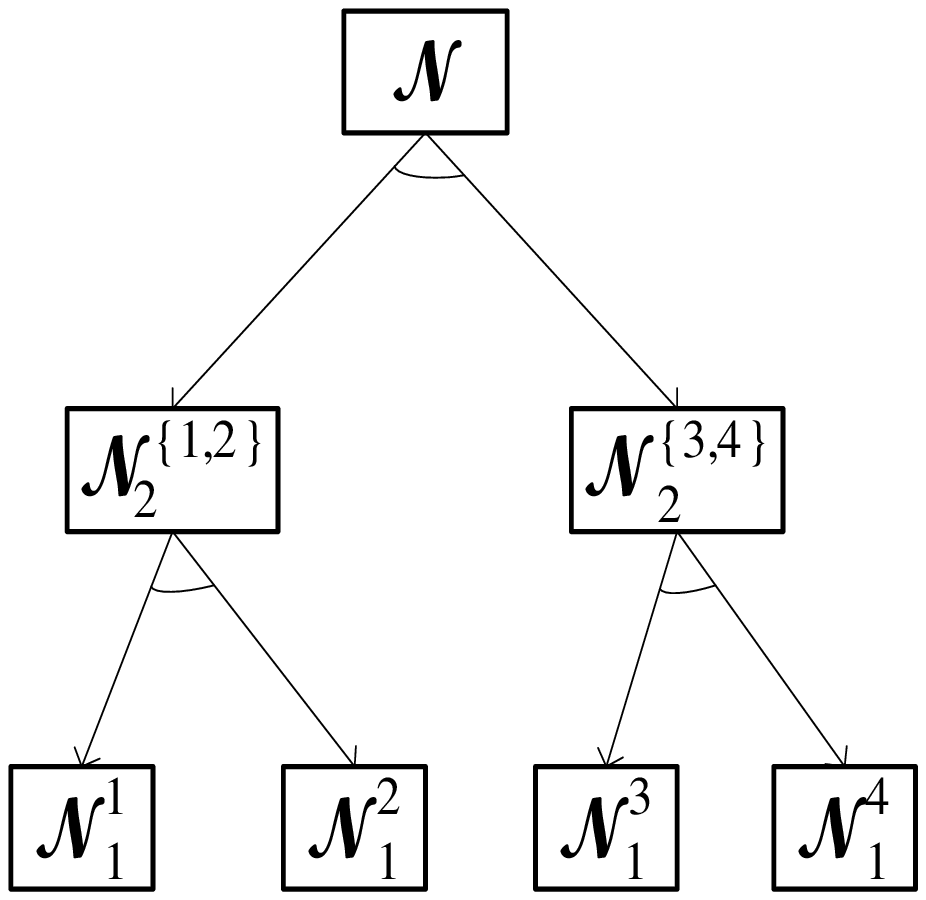,width=2.5cm} \label{Ch-NetworkedPartII-Fig-Tree1}
}
\subfigure[Plan 2]
{
    \psfig{file=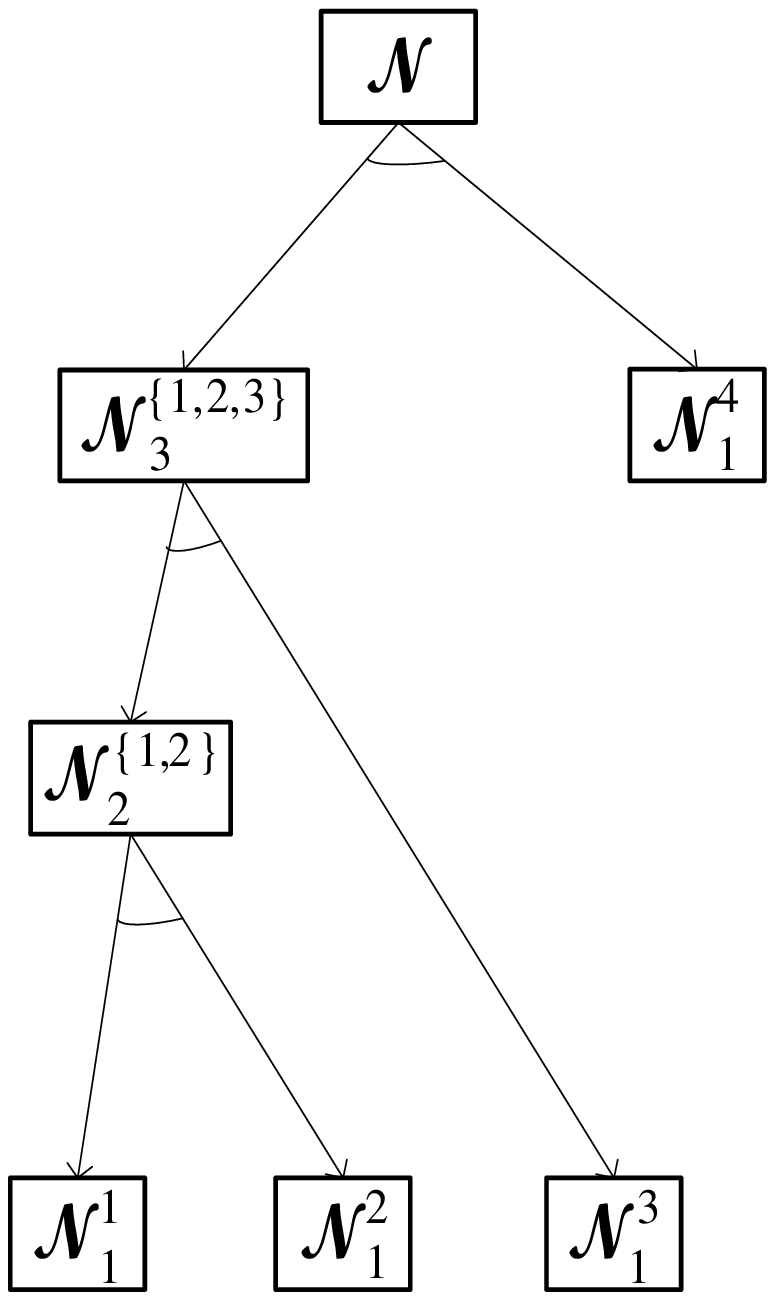,width=2.5cm} \label{Ch-NetworkedPartII-Fig-Tree2}
}
\subfigure[Plan 3]
{
    \psfig{file=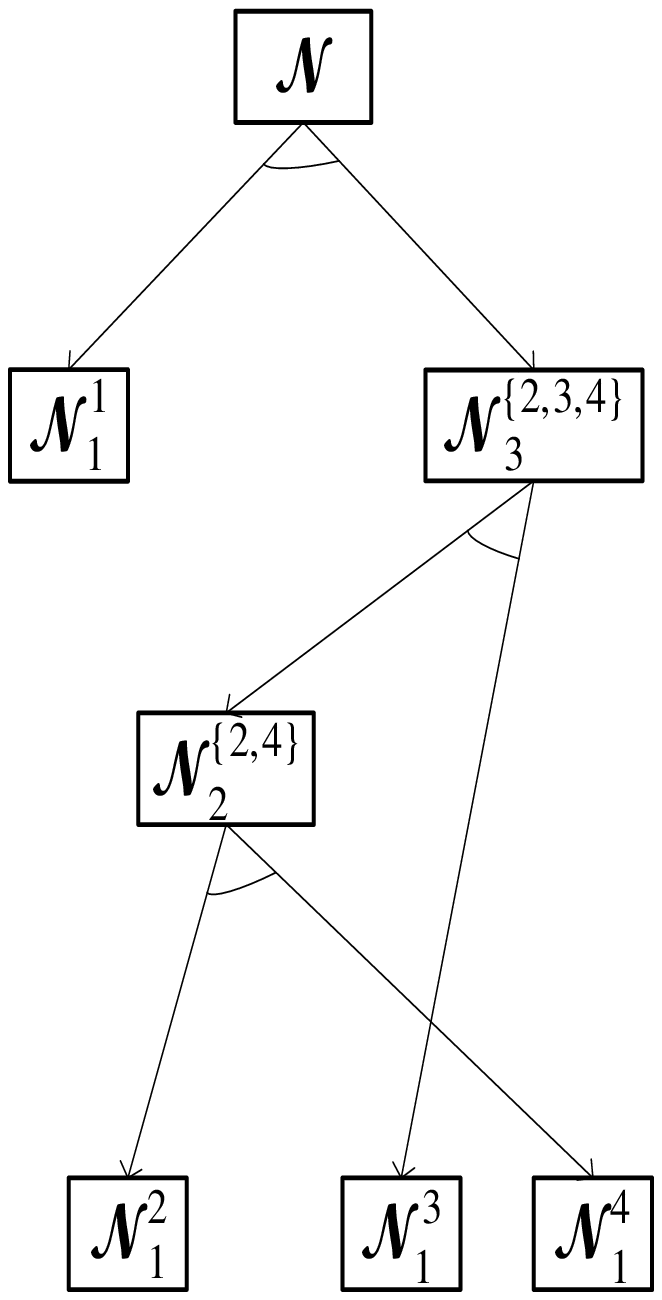,width=2.5cm} \label{Ch-NetworkedPartII-Fig-Tree3}
}
\subfigure[Fully generated AND/OR graph] {
    \psfig{file=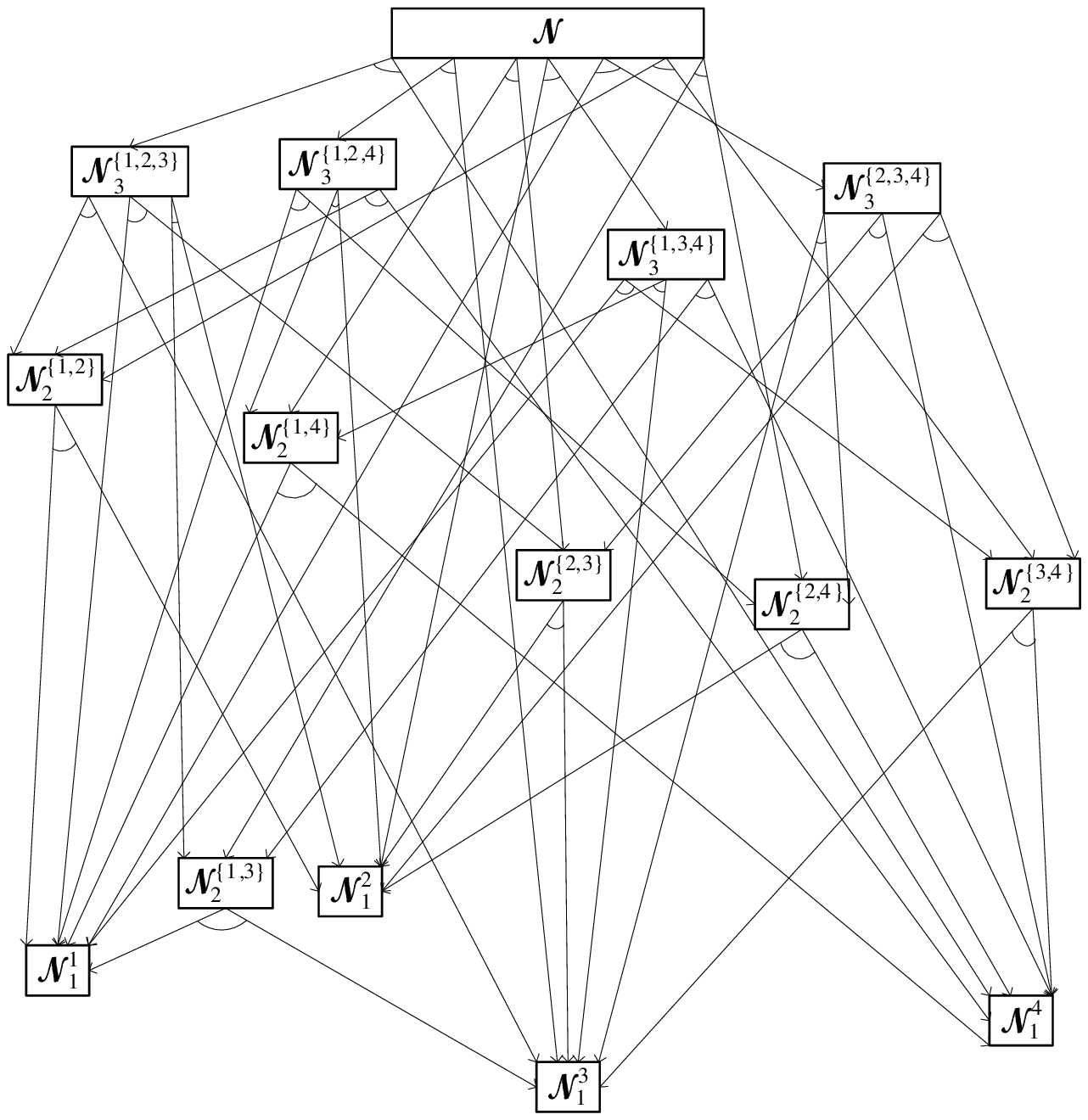,width=7cm} \label{Ch-NetworkedPartII-Fig-GeneratedANDORgraph}
}
\subfigure[Partially generated AND/OR graph] {
    \psfig{file=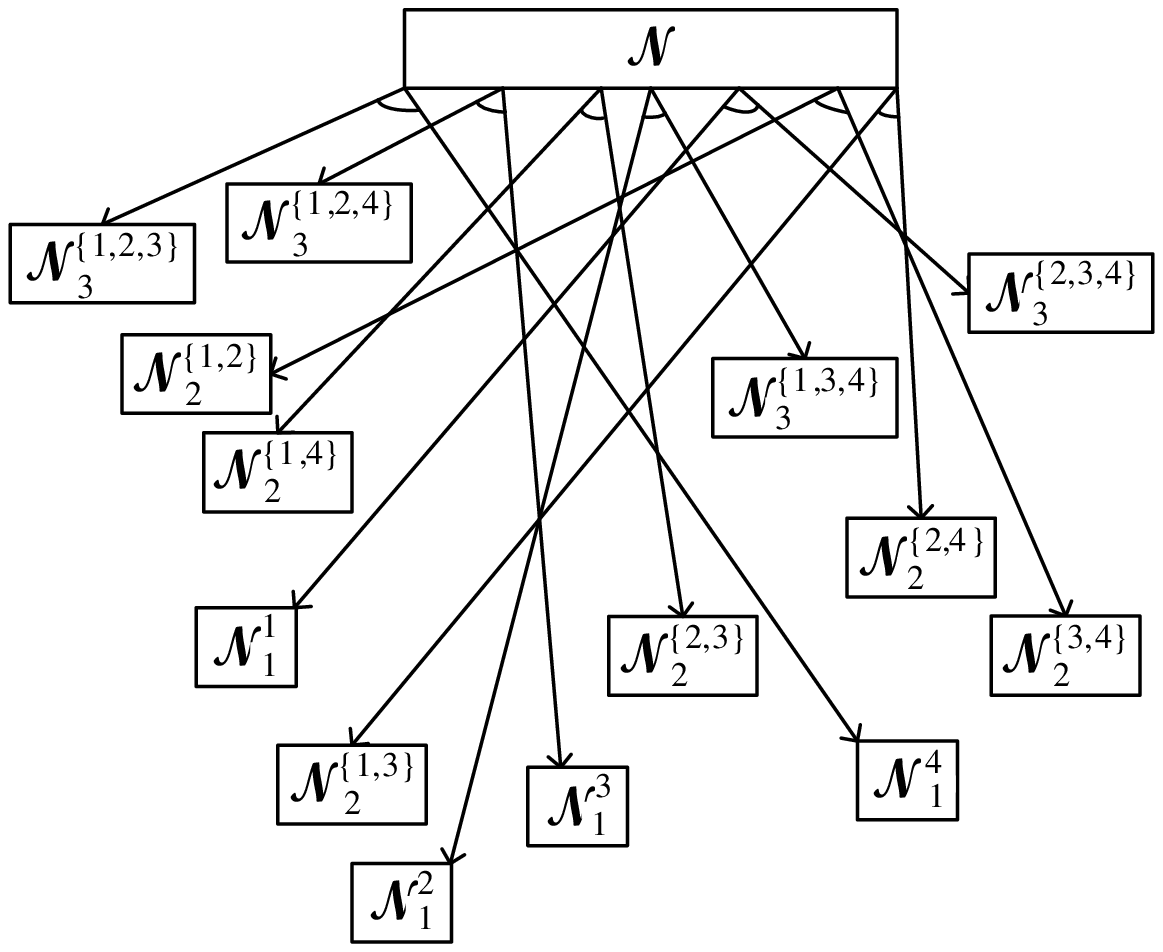,width=5cm} \label{Ch-NetworkedPartII-Fig-PartialANDORgraph}
}
\caption{The CRN, AND/OR graph of conflict resolution plans  and conflict resolution plans for a manufacturing transfer line system.}
\end{figure}

\begin{figure*}[!ht]
\centering
\subfigure
{
    \psfig{file=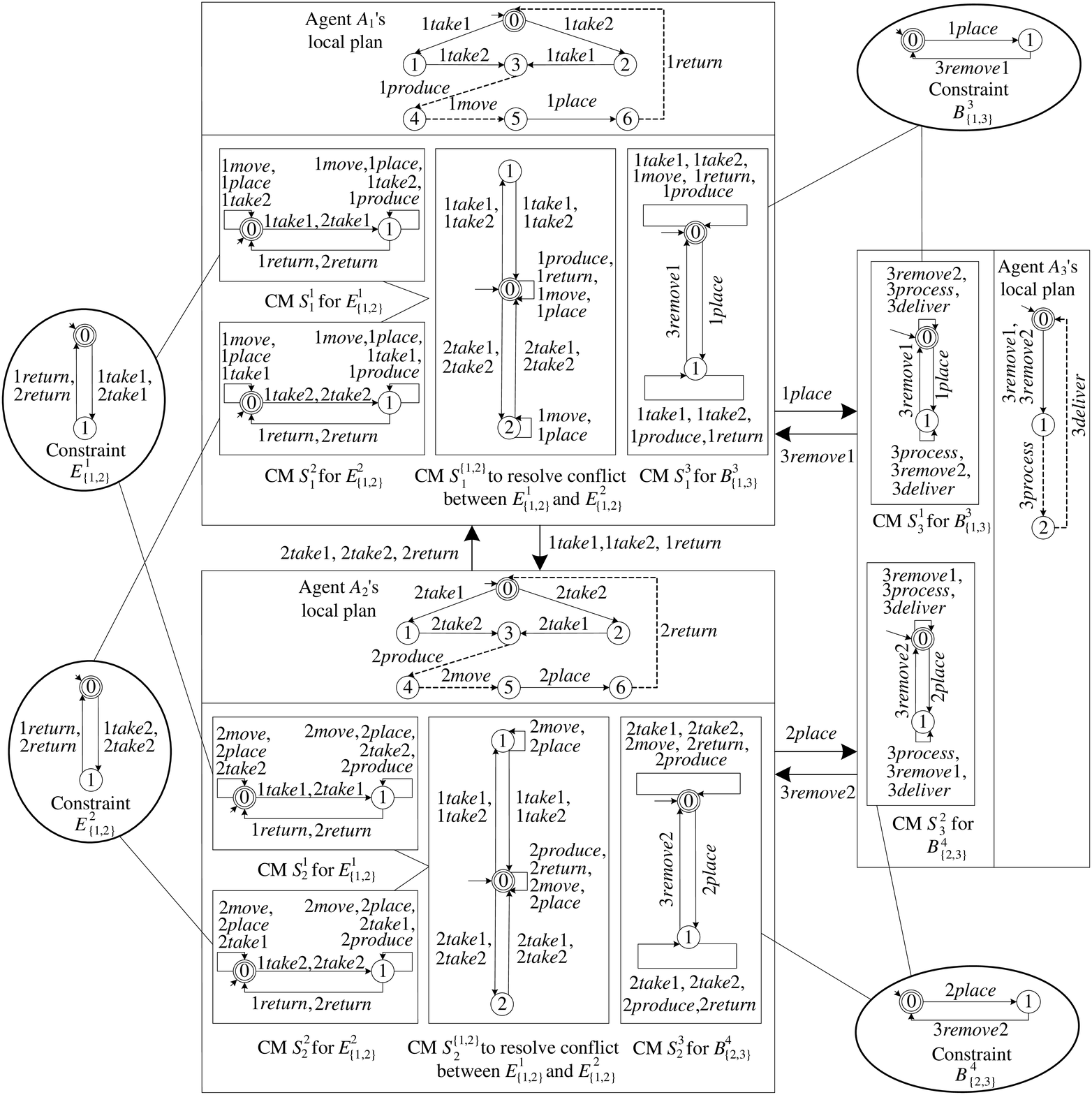,width=14cm}
}
\caption{Complete CM solution for a manufacturing transfer line system.} \label{Ch-NetworkedPartI-Fig-CompleteSolution}
\end{figure*}

\begin{example}
We now provide a solution for the manufacturing example. Following Step 1 of our approach presented, we use $CMBasicSubnet$ to design three local CM's for each of the agents $A_1$ and $A_2$, and two CM's for agent $A_3$. Each of these local CM's corresponds to a relevant constraint of the agents.

In Step 2.1, we need to generate a conflict resolution plan to completely and correctly composing together the subnets of the DCSN presented in Fig. \ref{Ch-NetworkedPartI-Fig-NW}. The CRN of this DCSN is shown in Fig. \ref{Ch-NetworkedPartII-Fig-CRN}. We apply $GenerateANDORGraph$ to decompose the CRN and generate the AND/OR graph plan shown in Fig. \ref{Ch-NetworkedPartII-Fig-GeneratedANDORgraph}. Each node in that graph represents a subnet of the DCSN. The root node represents the DCSN $\mathcal{N}$. There are six hyper-edges leaving that node, each of those represents one way the DCSN can be decomposed and points to the two nodes representing the resulting subnets. Similarly, the other nodes in the graph have a leaving hyper-edge for each possible way in which the subnets they represent can be decomposed. The AND/OR graph plans in Fig. \ref{Ch-NetworkedPartII-Fig-PartialANDORgraph}] are partially formed after the initial recursion, where all cut-sets for the CRN of DCSN $\mathcal{N}$ are computed \cite{B-NarsinghDeo1974} based on
fundamental cut-sets derived from a spanning tree highlighted over the CRN [Fig. \ref{Ch-NetworkedPartII-Fig-CRN}]. In every recursion,
each cut-set stored in $CutSets$ is a decomposition of the given subnet into a pair of subnets.

Figs.\ref{Ch-NetworkedPartII-Fig-Tree1}-\ref{Ch-NetworkedPartII-Fig-Tree3} show three conflict resolution plan trees that are extracted from the AND/OR graph. One important feature of the AND/OR graph tree representation of conflict resolution plans is that it shows explicitly the possibility of executing deconflicting operations in parallel. For example, while there are three deconflicting operations required by the conflict resolution plan represented by the tree in Fig. \ref{Ch-NetworkedPartII-Fig-Tree1}, the first two operations, for resolving the conflicts between $\mathcal{N}^1_1$ and $\mathcal{N}^2_1$, and $\mathcal{N}^3_1$ and  $\mathcal{N}^4_1$, can be performed simultaneously. Therefore, if there are two computational resources that can operate in parallel, the resolution plan can be completed in two sequential steps, with the first step to simultaneously deconflict between $\mathcal{N}^1_1$ and $\mathcal{N}^2_1$, and $\mathcal{N}^3_1$ and  $\mathcal{N}^4_1$, and the second step to deconflict between $\mathcal{N}^{\{1,2\}}_2$ and $\mathcal{N}^{\{3,4\}}_2$. In contrast, each of the other two trees in Figs. \ref{Ch-NetworkedPartII-Fig-Tree2}-\ref{Ch-NetworkedPartII-Fig-Tree3}  also has three operations. However, these operations have to be performed sequentially. Thus, no matter how many computational resources we have, each of these plans requires three sequential steps to complete.

To select an optimal conflict resolution plan from the generated AND/OR graph, we apply $HeuristicPlanSelection$ using the heuristic $H_p$, namely, one that allows maximal simultaneity in the execution of subnet composition operations. The selected conflict resolution plan is the one shown in Fig. \ref{Ch-NetworkedPartII-Fig-Tree1}. Following this plan and using $DeconfictBasicSubnet$ to compose subnets with conflict resolution, the complete solution is found and shown in Fig. \ref{Ch-NetworkedPartI-Fig-CompleteSolution}.
\end{example}

\section{Conclusion and Related Work} \label{Sect-RelatedWork-Conclusion}
This paper has introduced and addressed a novel multiagent coordination problem in a discrete-event formal languages and finite automata framework. The presented work is built on the results of \cite{CAAAI-Pham2008,CAAMAS-Seow2004,J-Seow2009,J-Pham2011}, generalizing the theory of multiagent coordination for a multi-constraint network of distributed agents.

Among related work under the same discrete-event paradigm, we have
earlier discussed the mathematical equivalence and conceptual
difference between our work on discrete-event multiagent
coordination and the well-established supervisory control of DES's
framework in our previous papers
\cite{CAAAI-Pham2008,CAAMAS-Seow2004,J-Seow2009,J-Pham2011}.
Elsewhere \cite{J-Pham2011, PhD-Pham2011}, we have also discussed
our discrete-event multiagent coordination framework in relation to
the distributed constraint satisfaction problem (DCSP)
\cite{B-MakotoYokoo2000}, multiagent planning
\cite{CAAAI-LukeHunsberger2002,CAAMAS-Planken2010,CAAMAS-Boerkoel2011,CAAMAS-JiayingShen2006}
and the Partially Observable Markov Decision Process (POMDP)
coordination framework \cite{J-ClaudiaGoldman2004}.

In a recent independent and emerging work
\cite{C-Cai2009,J-Cai2010,C-Cai2012a,C-Cai2012b}, a different
problem called supervisor localization is presented. For a DES $A$
consisting of $n \geq 2$ interacting local components $A_i$, $1 \leq
i \leq n$, with pair-wise disjoint event sets, the localization
problem focuses on decomposing (or localizing) a global supervisor
$S$ of $A$ into a set of local supervisors $\{S_i \mid 1 \leq i \leq
n \}$, with $S_i$ controlling $A_i$, while preserving the control
behavior of $S$ over $A$. Although communication minimization is not
explicitly considered in the supervisor localization solution, the
problem can be shown to be equivalent to our multiagent coordination
problem, i.e., Problem \ref{Ch-NetworkedPartI-Prob-Main}. However,
unlike the supervisor localization framework \cite{C-Cai2009}, our
multiagent framework clearly distinguishes the related but different
concepts of control and coordination by the Cartesian and
synchronous product operators \cite{CAAMAS-Seow2004}, respectively.
In distinguishing control and coordination, the mathematical
equivalence between coordination of localized supervisors and of
agents is established and discussed in \cite[Corollary
1]{J-Seow2009}. More importantly, in our opinion, this conceptual
difference brings into sharper focus the essence of our new
coordination problem, namely, designing built-in CM's - not
supervisors - for autonomous agents, and leads us to not prejudging
that the only means of CM synthesis is by first constructing
supervisors for a multiagent system. In addition, we note that the
intent of our framework is to naturally model active agents
coordinating through their CM's, whereas that of the framework
\cite{C-Cai2009} is apparently to model passive agents being
controlled by their interacting localized supervisors.

Finally, we note that the multiagent conflict resolution planning problem has not been addressed in the supervisor localization framework \cite{C-Cai2009,J-Cai2010,C-Cai2012a,C-Cai2012b} . In this paper, perhaps for the first time, we have proposed an efficient representation of conflict resolution plans for discrete-event agents using AND/OR graphs, and presented an algorithm to automatically generate an AND/OR graph representation of conflict resolution plans from a DCSN using cut-set theory \cite{B-NarsinghDeo1974}. Importantly, due to the mathematical equivalence between control and coordination, it is envisaged that our new results on multiagent conflict resolution planning can be adapted in the supervisor localization framework \cite{C-Cai2009,J-Cai2010,C-Cai2012a,C-Cai2012b} for systematic and efficient synthesis of localized supervisors.

\begin{small}
\section{Appendix}

\subsection{Proof of Lemma \ref{Ch-NetworkedPartI-Lemma-NonconflictTest2}} \label{Appendix2}
\begin{small}
%\begin{proof}
We have $SUP^h = Supcon(C^h_{J_h}, A_{J_h})$, $SUP^k = Supcon(C^k_{J_k}, A_{J_k})$ and $\Sigma^{\{h,k\}}_{CR} \supseteq \underset{i \in J_k \cap J_h}{\bigcup} \Sigma^{A_i}$. Suppose $P^h_{CR}$ is a $L_m(SUP^h)$-observer and $P^k_{CR}$ is a $L_m(SUP^k)$-observer. Then abstracting a theoretical result proved in \cite{CWODES-Pena2006}, it follows that $SUP^h \parallel SUP^k$ is nonblocking if and only if $P^h_{CR}(SUP^h) \parallel P^k_{CR}(SUP^k)$ is nonblocking. In other words, $\mathcal{N}^h_1$ and $\mathcal{N}^k_1$ are nonconflicting if and only if $P^h_{CR}(SUP^h) \parallel P^k_{CR}(SUP^k)$ is a nonblocking automaton.
%\end{proof}
\end{small}

\subsection{Proof of Lemma \ref{Ch-NetworkedPartI-Thm-LeiFeng}} \label{Appendix3}
\begin{small}
%\begin{proof}
Since the event sets of the agents in $\mathcal{N}^{\{h, k\}}_2$ are pair-wise disjoint, $SUP^h = Supcon(C^h_{J_h}, A_{J_h})$, $SUP^k = Supcon(C^k_{J_k}, A_{J_k})$ and $\Sigma^{\{h,k\}}_{CR} \supseteq \underset{i \in J_k \cap J_h}{\bigcup} \Sigma^{A_i}$, it follows from a theoretical result proved in \cite{J-LeiFeng2008} that if $P^h_{CR}$ is a $L_m(SUP^h)$-observer, $P^k_{CR}$ is a $L_m(SUP^k)$-observer and $(\forall i \in J_h \cup J_k)$ $P_{\Sigma^{A_i}, \Sigma^{\{h,k\}}_{CR}}$ is OCC for $L(A_i)$, then $Supcon[G,P^h_{CR}(SUP^h) \parallel P^k_{CR}(SUP^k)] \parallel SUP^h \parallel SUP^k$
$\equiv SUP^{\{h,k\}}$, namely,  $Supcon[G,P^h_{CR}(SUP^h) \parallel P^k_{CR}(SUP^k)]$ is a conflict resolution for $\mathcal{N}^h_1$ and $\mathcal{N}^k_1$.
%\end{proof}
\end{small}

\subsection{Proof of Lemma \ref{Ch-NetworkedPartI-Lemma-DeconflictBasicSubnet}} \label{Appendix4}
\begin{small}
%\begin{proof}
If the two basic subnets $\mathcal{N}^h_1$ and $\mathcal{N}^k_1$ are nonconflicting, the lemma is trivially true. Otherwise, by Step 3 of $DeconflictBasicSubnet$, for every agent $A_i$ with $\Sigma^{CR^{\{h,k\}}} \cap \Sigma^{A_i} \neq \emptyset$, we have $S^{\{h,k\}}_i=CMreduce(CR^{\{h,k\}}, A_i)$. Recall from \cite{J-Pham2011} that $CMreduce$ is a procedure that, given $CR^{\{h,k\}}$ and $A_i$, often returns a greatly state-size reduced CM automaton for agent $A_i$ achieving the same behavior of $A_i \parallel CR^{\{h,k\}}$. It follows that $\underset{\Sigma^{CR^{\{h,k\}}} \cap \Sigma^{A_i} \neq \emptyset}{\parallel} (A_i \parallel S^{\{h,k\}}_i) \equiv CR^{\{h,k\}}.$ For other agents that do no share events with $CR^{\{h,k\}}$, essentially no deconflicting CM is needed. Therefore, $\underset{i \in J_h \cup J_k}{\parallel} (A_i \parallel S^{\{h,k\}}_i) \equiv CR^{\{h,k\}}.$
%\end{proof}
\end{small}

\subsection{Proof of Theorem \ref{Ch-NetworkedPartI-Thm-Subnet}} \label{Appendix5}
\begin{small}
%\begin{proof}
If $\mathcal{N}^h_1$ and $\mathcal{N}^k_1$ are nonconflicting, the theorem is trivially true. Otherwise, by Lemma \ref{Ch-NetworkedPartI-Lemma-DeconflictBasicSubnet}, we have $\underset{i \in J_h \cup J_k}{\parallel} (A_i \parallel CM_i) \equiv (SUP^h \parallel SUP^k \parallel CR^{\{h,k\}}),$ where $CR^{\{h,k\}}$ is a conflict resolution for $\mathcal{N}^h_1$ and $\mathcal{N}^k_1$ computed in Step 2 of $DeconflictBasicSubnet$. Since, $(SUP^h \parallel SUP^k \parallel CR^{\{h,k\}}) \equiv SUP^{\{h,k\}}$, it follows that $\underset{i \in J_h \cup J_k}{\parallel} (A_i \parallel CM_i) \equiv SUP^{\{h,k\}}.$ Hence the theorem.
%\end{proof}
\end{small}

\subsection{Proof of Theorem \ref{Ch-NetworkedPartII-Prop-Heuristic}} \label{Appendix6}
\begin{small}
%\begin{proof}
Let the lowest $F$-value be $F^*$. By contradiction, assume that $HeuristicPlanSelection$ returns $tree$ with  $F(tree) > F^*$. Since $tree \in Trees(n_{root}, \Theta_{leaf})$, we also have $H(tree) = F(tree) > F^*$. Consider a partial tree $ptree$ that is a subgraph of an optimal plan $tree^* \in Trees(n_{root}, \Theta_{leaf})$ with $H(tree^*) = F(tree^*) = F^*$ and that is contained in $Q$ before $tree$ is extracted from $Q$ (there must always be such trees since an optimal solution always exists). Then, since $H$ is an admissible heuristic, we have $H(ptree) \leq F^*$.

We now have $H(ptree) \leq F^* < H(tree)$. Since in Step 3a, $HeuristicPlanSelection$ always extracts from $Q$ a tree with the lowest $H$-value, it follows that $tree$ will not be extracted from $Q$ before $ptree$ is. And when $ptree$ is extracted from $Q$, it will be expanded in Steps 3c and 3d, and eventually becomes $tree^*$ before $tree$ can ever be extracted from $Q$. The reason is that $H(tree^*) = F^* < H(tree)$ and $H$ is an admissible heuristic, meaning that any subgraph of $tree^*$ that is expanded from $ptree$ in Steps 3c and 3d will have its $H$-value smaller than that of $tree$ and therefore, extracted from $Q$ before $tree$. Finally, if $tree^*$ is ever be extracted from $Q$ in Step 3a, it will be returned as a solution by $HeuristicPlanSelection$. In other words, $tree$ will never be returned by $HeuristicPlanSelection$, contradicting our initial assumption. Hence the theorem.
%\end{proof}
\end{small}

\subsection{Proof of Lemma \ref{Ch-NetworkedPartII-Sect-LemmaAdmissible}} \label{Appendix7}
\begin{small}
%\begin{proof}
To prove this lemma, we have to prove that the following two conditions hold:

(i) If $tree \in Trees(n_{root}, \Theta_{leaf})$ then $H_p(tree) = F_p(tree)$, and (ii) $(\forall ptree \in Trees(n_{root},-))$ $(\forall tree \in Trees(n_{root},\Theta_{leaf}))$ ($ptree$ is a subgraph of $tree$) implies $H_p(ptree) \leq F_p(tree)$.

To prove (i), we shall show that if $tree \in Trees(n_{root}, \Theta_{leaf})$ then $H^\prime_p(tree) = Dept(tree)$. This can be done by a simple induction on the depth of trees as follows.

\begin{itemize}
  \item {\em Base:} First, since $\log_2(1) = 0$, any tree that contains only one node representing a basic subnet of $\mathcal{N}$ has both its $H^\prime_p$-value and its depth equal to 0.
  \item {\em Inductive Hypothesis:} Now, assume that any tree whose depth smaller than or equal to an integer $d \geq 0$ and whose terminal nodes are all in $\Theta_{leaf}$ has its depth equal to its $H^\prime_p$-value. We then show that any tree with depth $d + 1$ and with all terminal nodes in $\Theta_{leaf}$ will also have its depth equal to its $H^\prime_p$-value as follows.

      \begin{itemize}
        \item Let $tree = (n_I, h, tree_1, tree_2)$ be a tree with $Dept(tree) = d + 1$ and with every terminal node in $\Theta_{leaf}$. Since $Dept(tree) = 1 + max(Dept(tree_1), Dept(tree_2))$, $max(Dept(tree_1), Dept(tree_2)) = d$.
        \item It follows that both the depths of $tree_1$ and $tree_2$ are equal to or smaller than $d$. Furthermore, every terminal node of $tree_1$ and $tree_2$ is in $\Theta_{leaf}$. Therefore, by the inductive hypothesis, $Depth(tree_1) = H^\prime_p(tree_1)$ and $Depth(tree_2) = H^\prime_p(tree_2)$.
        \item It then follows that $max(Dept(tree_1), Dept(tree_2)) = max(H^\prime_p(tree_1), H^\prime_p(tree_2))$, or $Dept(tree) = 1 + max(H^\prime_p(tree_1), H^\prime_p(tree_2))$. By the definition of $H^\prime_p$, therefore, $Dept(tree) = H^\prime_p(tree)$.
      \end{itemize}

  \item Thus, by induction, if $tree \in Trees(n_{root}, \Theta_{leaf})$ then $H^\prime_p(tree) = Depth(tree)$. By the definitions of $F_p$ and $H_p$, it then follows that if $tree \in Trees(n_{root}, \Theta_{leaf})$, $H_p(tree) = F_p(tree)$.
\end{itemize}

To prove (ii), consider a partial tree $ptree$ in $T_\mathcal{N}$ that starts from $n_{root}$ and terminates at a set of nodes that are not necessarily leaf nodes. Consider a terminal node $n_t$ of $ptree$ that is not a leaf node, which represents a constraint-connected subnet of $\mathcal{N}$. Let $stree$ be an arbitrary tree that starts at $n_t$ and terminates at a subset of leaf nodes. $stree$ is then a sub-plan for $\mathcal{N}$, namely, a plan to synthesize the subnet represented by $n_t$. The depth of $stree$ must then be equal to or greater than $\log_2(NumBasicSubnet(n_t))$, since to synthesize the subnet represented by $n_t$, we need  to successively compose two different subnets of it at a time.

Since the depth of a tree starting from an arbitrary terminal node $n_t$ of $ptree$ and terminating at $\Theta_{leaf}$ is equal to or greater than $\log_2(NumBasicSubnet(n_t))$, by the recursive definitions of $H^\prime_p$, it follows that the depth of any tree in $Trees(n_{root}, \Theta_{leaf})$ that encompasses $ptree$ as a subgraph is equal to or greater than $H_p(ptree)$. In other words, $H_p$ is an admissible heuristic for $F_p$.
\end{small}
\end{small}

%\bibliography{ThesisRef}

\begin{thebibliography}{10}
\providecommand{\url}[1]{#1}
\csname url@rmstyle\endcsname
\providecommand{\newblock}{\relax}
\providecommand{\bibinfo}[2]{#2}
\providecommand\BIBentrySTDinterwordspacing{\spaceskip=0pt\relax}
\providecommand\BIBentryALTinterwordstretchfactor{4}
\providecommand\BIBentryALTinterwordspacing{\spaceskip=\fontdimen2\font plus
\BIBentryALTinterwordstretchfactor\fontdimen3\font minus
  \fontdimen4\font\relax}
\providecommand\BIBforeignlanguage[2]{{%
\expandafter\ifx\csname l@#1\endcsname\relax
\typeout{** WARNING: IEEEtran.bst: No hyphenation pattern has been}%
\typeout{** loaded for the language `#1'. Using the pattern for}%
\typeout{** the default language instead.}%
\else
\language=\csname l@#1\endcsname
\fi
#2}}

\bibitem{J-Seow2009}
K.~T. Seow, M.~T. Pham, C.~Ma, and M.~Yokoo, ``Coordination planning: Applying
  control synthesis methods for a class of distributed agents,'' \emph{IEEE
  Transactions on Control Systems Technology}, vol.~17, no.~2, pp. 405--415,
  2009.

\bibitem{J-PeterRamadge1987}
P.~J. Ramadge and W.~M. Wonham, ``Supervisory control of a class of discrete
  event processes,'' \emph{SIAM Journal of Control and Optimization}, 1987.

\bibitem{L-Wonham}
W.~M. Wonham, \emph{Notes on Control of Discrete-Event Systems ECE
  1636F/1637S}.\hskip 1em plus 0.5em minus 0.4em\relax Systems Control Group,
  University of Toronto, Updated 1st July 2012,
  http://www.control.toronto.edu/cgi-bin/dldes.cgi.

\bibitem{CAAAI-Pham2008}
M.~T. Pham and K.~T. Seow, ``Towards synthesizing optimal coordination modules
  for distributed agents,'' in \emph{Proceedings of the 23th AAAI Conference on
  Artificial Intelligence}, Chicago, Illinois, USA, July 2008, pp.
  1479--1480.

\bibitem{J-Pham2011}
------, ``Discrete-event coordination design for distributed agents,''
  \emph{IEEE Transactions on Automation Science and Engineering}, vol.~9,
  no.~1, pp. 70--82, 2012.

\bibitem{CAAMAS-Seow2004}
K.~T. Seow, C.~Ma, and M.~Yokoo, ``Multiagent planning as control synthesis,''
  in \emph{Proceedings of the 3rd International Joint Conference on Autonomous
  Agents and Multi-Agent Systems}, Columbia University, New York, July 2004, pp. 972--979.

\bibitem{B-Cassandras2008}
C.~G. Cassandras and S.~Lafortune, \emph{Introduction to Discrete Event
  Systems}.\hskip 1em plus 0.5em minus 0.4em\relax Springer, 2008.

\bibitem{B-NilsNilsson1980}
N.~J. Nilsson, \emph{Principles of artificial intelligence}.\hskip 1em plus
  0.5em minus 0.4em\relax New York: Springer-Verlag, 1980.

\bibitem{J-LeiFeng2008}
L.~Feng and W.~M. Wonham, ``Supervisory control architecture for discrete-event
  systems,'' \emph{IEEE Transactions on Automatic Control}, vol.~53, no.~6, pp.
  1449--1461, 2008.

\bibitem{B-NarsinghDeo1974}
N.~Deo, \emph{Graph Theory with Applications to Engineering and Computer
  Science}.\hskip 1em plus 0.5em minus 0.4em\relax New York: Prentice-Hall,
  1974.

\bibitem{B-StuartRussell2003}
S.~Russell and P.~Norvig, \emph{Artificial Intelligence: A Modern
  Approach}.\hskip 1em plus 0.5em minus 0.4em\relax Prentice Hall, 2003.

\bibitem{J-FengLin1988}
F.~Lin and W.~M. Wonham, ``On observability of discrete event systems,''
  \emph{Information Sciences}, vol.~44, no.~3, pp. 173--198, 1988.

\bibitem{SW-TCT}
W.~M. Wonham, \emph{Control Design Software: TCT}.\hskip 1em plus 0.5em minus
  0.4em\relax Developed by Systems Control Group, University of Toronto,
  Canada, Updated 1st July 2008,
  http://www.control.toronto.edu/cgi-bin/dlxptct.cgi.

\bibitem{J-KCWong2004}
K.~C. Wong and W.~M. Wonham, ``On the computation of observers in
  discrete-event systems,'' \emph{Discrete Event Dynamic Systems : Theory and
  Applications}, vol.~14, no.~1, pp. 55--107, 2004.

\bibitem{PhD-Pham2011}
M.~T. Pham, ``Discrete-event multiagent coordination: Framework and
  algorithms,'' School of Computer Engineering, Nanyang Technological
  University, PhD Thesis, 2011.

\bibitem{B-MakotoYokoo2000}
M.~Yokoo, \emph{Distributed Constraint Satisfaction : Foundations of
  Cooperation in Multi-Agent Systems}.\hskip 1em plus 0.5em minus 0.4em\relax
  Springer-Verlag, Heidelberg, Germany, 2000, \ Springer Series on Agent
  Technology.

\bibitem{CAAAI-LukeHunsberger2002}
L.~Hunsberger, ``Algorithms for a temporal decoupling problem in multi-agent
  planning,'' in \emph{Proceedings the 18th National Conference on Artificial
  Intelligence}, Edmonton, Alberta, July-August 2002, pp.
  468--475.

\bibitem{CAAMAS-Planken2010}
L.~Planken, M.~de~Weerdt, and C.~Witteveen, ``Optimal temporal decoupling in
  multiagent systems,'' in \emph{Proceedings of the 9th International Joint
  Conference on Autonomous Agents and Multi-Agent Systems},
  Toronto, May 2010, pp. 789--796.

\bibitem{CAAMAS-Boerkoel2011}
J.~C.~B. Jr. and E.~H. Durfee, ``Distributed algorithms for solving the
  multiagent temporal decoupling problem,'' in \emph{Proceedings of the 10th
  International Joint Conference on Autonomous Agents and Multi-Agent Systems},
  Taipei, May 2011, pp. 141--148.

\bibitem{CAAMAS-JiayingShen2006}
J.~Shen and V.~R. Lesser, ``Communication management using abstraction in
  distributed bayesian networks,'' in \emph{Proceedings of the Fifth
  International Joint Conference on Autonomous Agents and Multi-Agent Systems},
  Future University-Hakodate, Hakodate, May 2006, pp.
  622--629.

%\bibitem{CAAAI-Georgeff1983}
%M.~Georgeff, ``Communication and interaction in multiagent planning,'' in
%  \emph{Proceedings the 3rd National Conference on Artificial Intelligence},
%  Washington, D.C, July 1983, pp. 125--129.

\bibitem{J-ClaudiaGoldman2004}
C.~V. Goldman and S.~Zilberstein, ``Decentralized control of cooperative
  systems: Categorization and complexity analysis,'' \emph{Journal of
  Artificial Intelligence Research}, vol.~22, pp. 143--174, 2004.

\bibitem{C-Cai2009}
K.~Cai and W.~M. Wonham, ``Supervisor localization: A top-down approach to
  distributed control of discrete-event systems,'' in \emph{Second
  Mediterranean Conference on Intelligent Systems and Automation (CISA'09). AIP
  Conference Proceedings}, vol. 1107, Zarzis, March 2009, pp.
  302--308.

\bibitem{J-Cai2010}
------, ``Supervisor localization: a top-down approach to distributed control
  of discrete-event systems,'' \emph{IEEE Transactions on Automatic Control},
  vol.~55, no.~3, pp. 605--618, 2010.

\bibitem{C-Cai2012a}
------, ``New results on supervisor localization, with application to
  multi-agent formations,'' in \emph{Proceeding of Workshop of Discrete-Event
  Systems}, Guadalajara, Mexico, October 2012, pp. 233--238.

\bibitem{C-Cai2012b}
------, ``Supervisor localization of discrete-event systems based on state tree
  structures,'' in \emph{Proceeding of the 51st IEEE Conference on Decision and
  Control}, Maui, Hawaii, December 2012, pp. 5822--5827.

\bibitem{C-Queiroz2000}
M.~de~Queiroz and J.~Cury, ``Modular control of composed systems,'' in
  \emph{Proceedings of the American Control Conference}, Chicago, IL, June
  2000, pp. 4051--4055.

\bibitem{CWODES-Pena2006}
P.~N. Pena, J.~E.~R. Cury, and S.~Lafortune, ``Testing modularity of local
  supervisors: An approach based on abstractions,'' in \emph{Proceedings of the
  8th International Workshop on Discrete Event Systems}, Ann Arbor,
  Michigan, July 2006, pp. 107--112.

\bibitem{J-RongSu2004}
R. Su and W.~M. Wonham, ``Supervisor reduction for discrete-event systems,'' \emph
{Discrete Event Dynamic Systems : Theory and Applications}, vol.~14, no.~1, pp. 31--53, 2004.

\end{thebibliography}
%\bibliographystyle{plain}
%\bibliographystyle{IEEEtranS}

\end{document}